\documentclass[12pt]{article}

\usepackage{amssymb}
\usepackage[dvips]{graphicx}

\def\appendix#1{
  \addtocounter{section}{1}
  \setcounter{equation}{0}
  \renewcommand{\thesection}{\Alph{section}}
  \section*{Appendix \thesection\protect\indent \parbox[t]{11.715cm} {#1}}
  \addcontentsline{toc}{section}{Appendix \thesection\ \ \ #1}
  }

\def\l{\Big(}
\def\r{\Big)}

\newcommand {\bd}{\begin{displaymath}}
\newcommand {\ed}{\end{displaymath}}
\newcommand {\eq}{\begin{equation}}
\newcommand {\beq}{\begin{equation}}
\newcommand {\eeq}{\end{equation}}
\newcommand {\beqa}{\begin{eqnarray}}
\newcommand {\eeqa}{\end{eqnarray}}
\newcommand {\n}{\nonumber \\}
\newcommand {\tr}{{\rm tr\,}}
\newcommand {\Tr}{\mbox{Tr\,}}

\newcommand {\ee}{\mbox{e}}

\newcommand {\dd}{\mbox{d}}

\newcommand {\del}{\partial}

\newcommand {\defeq}{\stackrel{\rm def}{=}}
\font\mybbs=msbm10 at 9pt
\def\bbs#1{\hbox{\mybbs#1}}
\font\mybb=msbm10 at 12pt
\def\bb#1{\hbox{\mybb#1}}

\def\IT{{\bb T}}
\def\ITs{{\bbs T}}
\def\IC{{\bb C}}
\def\IR{{\bb R}}
\def\IZ{{\bb Z}}

\def\IZs{{\bbs Z}}

\newcommand{\id}{{1\!\!1}} 

\begin{document}

\setlength{\oddsidemargin}{0cm}
\setlength{\baselineskip}{7mm}

\begin{titlepage}

\baselineskip=14pt

 \renewcommand{\thefootnote}{\fnsymbol{footnote}}
\begin{normalsize}
\begin{flushright}
\begin{tabular}{l}
NBI-HE-00-21\\
ITEP-TH-11/00\\
hep-th/0004147\\
\hfill{ }\\
April 2000
\end{tabular}
\end{flushright}
  \end{normalsize}

    \begin{Large}
       \begin{center}
{Lattice Gauge Fields and\\ Discrete Noncommutative Yang-Mills Theory}\\

       \end{center}
    \end{Large}

\vspace{1cm}

\begin{center}
           J. A{\sc mbj\o rn}$^{1)}$\footnote
            {
e-mail address :
ambjorn@nbi.dk},
           Y.M. M{\sc akeenko}$^{1)\,2)}$\footnote
            {
e-mail address :
makeenko@nbi.dk},
           J. N{\sc ishimura}$^{1)}$\footnote{
Permanent address : Department of Physics, Nagoya University,
Nagoya 464-8602, Japan\\
e-mail address : nisimura@nbi.dk}
           {\sc and}
           R.J. S{\sc zabo}$^{1)}$\footnote
           {e-mail address : szabo@nbi.dk}\\
      \vspace{1cm}
        $^{1)}$ {\it The Niels Bohr Institute\\ Blegdamsvej 17, DK-2100
                 Copenhagen \O, Denmark}\\[4mm]
        $^{2)}$ {\it Institute of Theoretical and Experimental Physics}\\
               {\it B. Cheremushkinskaya 25, 117218 Moscow, Russia} \\
\end{center}

\hspace{5cm}

\begin{abstract}
\noindent

We present a lattice formulation of noncommutative Yang-Mills theory in
arbitrary even dimensionality. The UV/IR mixing characteristic of
noncommutative field theories is demonstrated at a completely nonperturbative
level. We prove a discrete Morita equivalence between ordinary Yang-Mills
theory with multi-valued gauge fields and noncommutative Yang-Mills theory with
periodic gauge fields. Using this equivalence, we show that generic
noncommutative gauge theories in the continuum can be regularized
nonperturbatively by means of {\it ordinary} lattice gauge theory with 't~Hooft
flux. In the case of irrational noncommutativity parameters, the rank of the
gauge group of the commutative lattice theory must be sent to infinity in the
continuum limit. As a special case, the construction includes the recent
description of noncommutative Yang-Mills theories using twisted large $N$
reduced models. We study the coupling of noncommutative gauge fields to matter
fields in the fundamental representation of the gauge group using the lattice
formalism. The large mass expansion is used to describe the physical meaning of
Wilson loops in noncommutative gauge theories. We also demonstrate Morita
equivalence in the presence of fundamental matter fields and use this property
to comment on the calculation of the beta-function in noncommutative quantum
electrodynamics.

\end{abstract}
\vfill
\end{titlepage}
\vfil\eject

\setcounter{page}{2}

\tableofcontents

\renewcommand{\thefootnote}{\arabic{footnote}}

\baselineskip=18pt

\setcounter{footnote}{0}
\section{Introduction and summary}
\setcounter{equation}{0}

\renewcommand{\thefootnote}{\arabic{footnote}}

The foundations of noncommutative geometry \cite{NCgeom,FGR} are inspired in
large part by the fundamental principles of quantum mechanics.
Just as in quantum mechanics physical observables are replaced with operators,
and thereby satisfy the Heisenberg uncertainty principle,
in noncommutative geometry spacetime coordinates are replaced with noncommuting
operators. The corresponding smearing of spacetime coordinates in this way fits
nicely into the ideas behind spacetime uncertainty relations \cite{yoneya} and
the concept of minimum length in string theory \cite{minimal}. This heuristic
correspondence has been used to suggest that noncommutative geometry provides a
natural framework to describe nonperturbative aspects of string theory
\cite{FGR,NCstrings}. This belief is further supported by the fact that
Matrix Theory \cite{BFSS} and the IIB matrix model \cite{IKKT},
which are conjectured to provide nonperturbative definitions of string
theories, give rise to noncommutative Yang-Mills theory on toroidal
compactifications \cite{CDS}. The particular noncommutative toroidal
compactification is interpreted as being the result of the presence of a
background Neveu-Schwarz two-form field, and it can also be understood in the
context of open string quantization in D-brane backgrounds \cite{MD,SW}.
Furthermore, in Ref.~\cite{AIIKKT} it has been shown that the IIB matrix model
with D-brane backgrounds is described by noncommutative Yang-Mills theory.

The early motivation \cite{snyder} for studying quantum field theory on
noncommutative spacetimes was that, because of the spacetime uncertainty
relation, the introduction of noncommutativity would provide a natural
ultraviolet regularization. However, more recent perturbative calculations
\cite{filk}--\cite{MRS} have shown that planar noncommutative Feynman diagrams
contain exactly the same ultraviolet divergences that their commutative
counterparts do, which implies that the noncommutativity does not serve as an
ultraviolet regulator.
One therefore needs to introduce some other form of regularization
to study the dynamics of noncommutative field theories.
On the other hand, it has been found that the ultraviolet divergences in
non-planar Feynman diagrams \cite{MRS,UVIR} exhibit an intriguing mixing of
ultraviolet and infrared scales, which can also be described using
string-theoretical approaches \cite{BCR,IIKK1}. Heuristically, this UV/IR
mixing can be understood in terms of the induced uncertainty relations among
the spacetime coordinates. If one measures a given spacetime coordinate with
some high precision, then the remaining spacetime directions will generally
extend because of the smearing. Furthermore, noncommutative solitons which do
not have counterparts in ordinary field theory have been discovered \cite{GMS}
for sufficiently large values of the noncommutativity parameters, and it has
also been shown \cite{IIKK1} that noncommutative Yang-Mills theory in four
dimensions naturally includes gravity.

In order to investigate further the non-trivial dynamics of noncommutative
field theories, it is important therefore to develop a nonperturbative
regularization of these theories. Such a program has been put forward in
Refs.~\cite{AIIKKT,IIKK,IIKK1},\cite{BM}--\cite{AMNS2} and it is similar to
earlier works \cite{earlier} based on the mapping between large $N$ matrices
and spacetime fields.
In particular, in Ref.~\cite{AMNS} a unified framework was presented
which naturally interpolates between the two ways that
noncommutative Yang-Mills theory has appeared in the context of matrix model
formulations of string theory, namely the compactification of Matrix theory
and the twisted large $N$ reduced model. The model proposed was a finite $N$
matrix model defined by the twisted Eguchi-Kawai model \cite{EK,GO}
with a quotient condition analogous to the ones considered in
Refs.~\cite{CDS,cmp}. It was interpreted as a lattice formulation of
noncommutative gauge theory with manifest star-gauge invariance.
The formulation naturally includes Wilson lattice gauge theory \cite{Wilson}
for a particular choice of parameters even at finite $N$. In Ref.~\cite{AMNS2},
the lattice formulation was reconsidered from a general point of view without
specifying any representation of the noncommutative algebra generated by the
spacetime coordinates. This enabled an extension of the formalism to arbitrary
even dimensions and also to allow for the minimal coupling to matter fields in
the fundamental representation of the gauge group. Furthermore, in this
approach one produces finite dimensional representations of the noncommutative
geometry in much the same spirit as Weyl's finite version of quantum mechanics
\cite{weyl}. Discrete versions of noncommutative geometry using random lattices
and graphs, as well as their applications to finite dimensional versions of
noncommutative gauge theories, have also been studied in Ref.~\cite{NClattice}.

In this paper we will show further that noncommutative Yang-Mills theory in an
arbitrary even number of spacetime dimensions, with either rational or
irrational valued (dimensionless) noncommutativity parameters and with gauge
fields of arbitrary topological charge, can be regularized nonperturbatively by
using {\em commutative} lattice gauge theories with twisted boundary conditions
on the gauge fields \cite{cmp,tHooft}. We will do so by presenting an extensive
description of the lattice formulation of noncommutative gauge theories which
was reported in Ref.~\cite{AMNS2}. The first striking fact we will find in the
general lattice formulation of noncommutative field theories is that the
discretization of spacetime inevitably requires that it be compactified, in
order to satisfy certain consistency conditions on the noncommutative algebra
generated by the spacetime coordinates. It then follows that the lattice
spacing must be much smaller than the length scale determined by the
noncommutativity parameters, and so the continuum limit does not commute with
the commutative limit. This demonstates the UV/IR mixing mentioned above at a
completely nonperturbative level. We will also show that the dimensionless
noncommutativity parameters are necessarily rational-valued on the lattice,
which leads immediately to the possibility of obtaining finite dimensional
representations of the noncommutative algebra as alluded to above. Irrational
noncommutativity parameters can then be obtained by taking the continuum limit
appropriately.

Using the lattice formulation, we can regularize noncommutative Yang-Mills
theories with periodic boundary conditions on the gauge fields, arbitrary
noncommutativity parameters and arbitrary spacetime periodicities. We will then
demonstrate on the lattice the Morita equivalence \cite{Morita} between
ordinary, commutative Yang-Mills theory with twisted boundary conditions on the
gauge fields
and noncommutative Yang-Mills theory with periodic gauge fields.
Using this equivalence, we will find that noncommutative Yang-Mills theory
with single-valued gauge fields can actually be regularized by means of
{\em commutative} lattice gauge theory with background magnetic flux.
In order to obtain a continuum field theory with irrational-valued
dimensionless noncommutativity parameters, the rank of the gauge group of the
regulating commutative theory must be taken to infinity as one takes the
continuum limit.
As a special case, the construction includes the recent proposal of using
twisted large $N$ reduced models \cite{AIIKKT} for a concrete definition of
noncommutative gauge theory. We shall find, however, that the class of
noncommutative Yang-Mills theories that one can obtain using twisted large $N$
reduced models is not the most general one.

The formalism also allows for the coupling of gauge fields to matter fields
in the fundamental representation of the gauge group. We will carry out a large
mass expansion of the matter-coupled theory and clarify the physical meaning of
Wilson loops in noncommutative gauge theories
\cite{IIKK,AMNS,AMNS2,NCloops,Alekseev}. A remarkable property of these Wilson
loops \cite{IIKK} is that star-gauge invariance does not require that
the loop be closed, but it does require that the relative distance vector
between the two ends of the loop be proportional to its total momentum.
Namely, as one increases the total momentum, the loop extends in spacetime.
This is another manifestation of UV/IR mixing due to the noncommutativity.
We will demonstrate how noncommutative Wilson loops
naturally arise from matter field averages in
noncommutative gauge theory. By calculating the effective action induced by
matter and star-gauge invariant observables constructed out of the matter
fields,
we shall see that noncommutative Wilson loops
play a very fundamental role, just like ordinary Wilson loops
do in commutative gauge theories \cite{Wilson}.
We will also show explicitly how star-gauge invariant observables
reduce smoothly to ordinary gauge invariant observables
in the commutative limit for fixed gauge backgrounds.

Morita equivalence in the presence of fundamental matter fields is also proven.
As a special case, we obtain the twisted Eguchi-Kawai model with fundamental
matter which was introduced in Ref.~\cite{Das} as a model which
reproduces ordinary large $N$ gauge theory in the Veneziano limit.
These Morita equivalences clarify the interpretations of various quantities in
noncommutative gauge theory. For instance, it is known that only planar
noncommutative Feynman diagrams contribute to the beta-function at one-loop
order \cite{ren,MRS,Hayakawa}, and in the case of noncommutative quantum
electrodynamics it is given by \cite{Hayakawa}
\beq
\beta(g^2)=-\frac{g^4}{4\pi^2}\,\left(\frac{11}3-\frac23\,n_f\right) \ ,
\label{1loopbeta}\eeq
where $n_f$ is the number of fermion flavours. The beta-function
(\ref{1loopbeta}) coincides with that of ordinary ${\rm U}(N)$
Yang-Mills theory coupled to
$N_f=n_fN$ flavours of fermion fields in the large $N$ (planar) limit (after an
appropriate rescaling of the Yang-Mills coupling constant $g$). This
coincidence can be understood as a consequence of the Morita equivalence that
we shall derive. In fact, the results of such calculations indicate that the
phenomenon of Morita equivalence in noncommutative gauge theories holds beyond
the classical level in regularized perturbation theory. We will further show
that the star-gauge invariant observables of two Morita equivalent
noncommutative Yang-Mills theories are in a one-to-one correspondence. Owing to
this fact, we can define regularized
correlation functions of star-gauge invariant observables
in noncommutative Yang-Mills theory with multi-valued gauge fields
by using the lattice regularization of the corresponding Morita equivalent
noncommutative gauge theory with periodic gauge fields. This Morita equivalence
property will be the key feature which permits the discretization of generic
noncommutative Yang-Mills theories in the continuum.

The organization of the remainder of this paper is as follows.
In Section \ref{review} we present a pedagogical introduction to
noncommutative field theory, and in particular noncommutative
Yang-Mills theory, for the sake of completeness and
in order to set up the notations to be used throughout the paper.
In Section \ref{Morita}, we present a detailed and very general field
theoretical derivation of the Morita equivalence relation between
noncommutative gauge theories and demonstrate the one-to-one correspondence
between star-gauge invariant observables in the Morita equivalent theories.
In Section \ref{lattice}, we construct the discrete version of noncommutative
field theory. We show that the lattice formulation exhausts the
noncommutative Yang-Mills theories in arbitrary even dimensions,
given the Morita equivalence properties described in Section \ref{Morita}.
In Section \ref{explicit} we establish Morita equivalence
on the lattice, which allows us to regularize noncommutative Yang-Mills
theories by means of commutative lattice gauge theories with twisted boundary
conditions on the gauge fields. We also describe the relationship to finite
dimensional matrix model representations of noncommutative gauge theories.
Finally, in Section \ref{hopping} we introduce matter fields in the fundamental
representation of the gauge group and study the properties of star-gauge
invariant observables as well as Morita equivalence in the presence of matter
fields.

\section{Quantum field theory on noncommutative spaces}
\label{review}
\setcounter{equation}{0}

In this Section we will briefly review some aspects of constructing
noncommutative gauge theories in the continuum, in a way that will enable us to
later construct a lattice version of the field theory. We will start with a
simple scalar field theory to introduce the ideas, and then move on to a
description of noncommutative Yang-Mills theory and its observables.

\subsection{Scalar field theory}
\label{scalar}

Consider the scalar quantum field theory which is described by the partition
function
\beqa
Z&=& \int \it {\cal D} \phi ~ \ee ^{-S[\phi]} \n
S&=& \int \dd ^D x
\left( \frac{1}{2}\,\Bigl(\del _\mu \phi\Bigr)^2
+ \frac{1}{2}\,\phi ^2
+ \frac{1}{4 \mbox{!}}\,\phi ^4 \right) \ ,
\label{scalarft}\eeqa
where $\phi (x)$ is a real-valued scalar field on $D$-dimensional Euclidean
spacetime $\IR^D$. For fields in a Schwartz space of functions of sufficiently
rapid decrease at infinity, we may use the Fourier transformation
\beq
\phi (x) = \int \frac{\dd ^D k}{(2\pi)^D} ~
 \tilde{\phi} (k)~\ee ^ {i k_\mu  x_\mu} \ ,
\label{Fourier}
\eeq
with $\tilde\phi(-k)=\tilde\phi(k)^*$. The first step in generalizing a quantum
field theory on an ordinary spacetime to one on a noncommutative spacetime is
to replace the local coordinates $x_\mu$ by
hermitian operators $\hat{x}_\mu$ obeying the commutation relations
\beq
[\hat{x}_\mu,\hat{x}_\nu] = i\,\theta_{\mu\nu} \ ,
\label{Xcommutation}
\eeq
where $\theta_{\mu\nu}=-\theta_{\nu\mu}$ are real-valued
c-numbers with dimensions of length squared.
Consequently, fields on spacetime are replaced by operators.
Replacing $x_\mu$ in (\ref{Fourier}) by $\hat{x}_\mu$, we obtain
\beq
\hat{\phi} =\int \frac{\dd ^D k}{(2\pi)^D} ~\tilde{\phi} (k)~
 \ee ^ {i k_\mu \hat{x}_\mu} \ .
\label{NCPhi}
\eeq
The operator-ordering ambiguity which exists when we replace
$x_\mu$ by $\hat{x}_\mu$ is fixed in (\ref{NCPhi}) by requiring
covariance and hermiticity of the operator $\hat{\phi}$. The operator
(\ref{NCPhi}) is called a Weyl operator or the Weyl symbol of the field
$\phi(x)$, and this method of constructing a noncommutative field theory is
sometimes refered to as Weyl quantization.

Combining (\ref{Fourier}) and (\ref{NCPhi}),
one can obtain an explicit map $\hat{\Delta} (x)$
which transforms a field $\phi (x)$ to an operator $\hat{\phi}$ as
\beqa
\label{maptophi}
\hat{\phi} &=& \int \dd ^ D x ~ \phi (x)\,\hat{\Delta} (x) \\
\hat{\Delta} (x) &=&
\int\frac{\dd ^D k}{(2\pi)^D} ~
\ee ^ {i k_\mu \hat{x}_\mu}
{}~\ee ^{-i k_\mu x_\mu} \ .
\label{map}
\eeqa
This means that the field $\phi(x)$ can be thought of as the coordinate space
representation of the operator $\hat\phi$. Note that the operator (\ref{map})
is hermitian, $\hat{\Delta} (x) ^\dag =\hat{\Delta} (x)$, and in the
commutative case $\theta_{\mu\nu}=0$ it reduces to a delta-function
$\delta^D(\hat x-x)$. However, for $\theta_{\mu\nu}\neq0$, $\hat\Delta(x)$ is a
complicated map. We may also introduce an anti-hermitian derivation $\hat{\del}
_\mu$ through the
commutation relations
\beq
\left[\hat{\del}_\mu\,,\, \hat{x} _\nu \right]
= \delta _ {\mu\nu}~~~~~~,~~~~~~\left[\hat\del_\mu\,,\,\hat\del_\nu
\right]=i\,c_{\mu\nu} \ ,
\eeq
where $c_{\mu\nu}$ are real-valued c-numbers.\footnote{The c-numbers
$c_{\mu\nu}$ turn out to be irrelevant for our purposes, since the operator
$\hat{\del}_\mu$ will only appear in commutator brackets.
For this reason it is conventional to set $c_{\mu\nu}=0$.
However, this restriction is not necessary, and
indeed we will encounter cases with non-vanishing $c_{\mu\nu}$ later on.}
One can show that
\beqa
\left[ \hat{\del}_\mu\,,\, \hat{\phi} \right]
&=& \int \dd ^D x ~
\del _\mu \phi (x)\,\hat{\Delta} (x)  \n
\left[ \hat{\del}_\mu\,,\, \hat{\Delta}(x) \right]
&=& - \del _\mu \hat{\Delta}(x) \ .
\label{derivaction}\eeqa

{}From (\ref{derivaction}) it follows that any translation generator can be
represented by the operator
$\ee ^{v_\mu \hat{\del}_\mu}$, $v _\mu \in \IR$, which satisfies
\beq
\ee ^{v_\mu \hat{\del}_\mu }~\hat{\Delta} (x)~\ee^{-v_\mu \hat\del _\mu}
=\hat{\Delta} (x+v) \ .
\eeq
The existence of such an operator implies that $\Tr \hat{\Delta} (x)$ is
independent of $x$ for any trace Tr on the algebra of operators.
Therefore, one can represent an integration of fields
on the spacetime as
\beq
\Tr \hat{\phi}  = \int \dd ^D x ~ \phi (x) \ ,
\label{integration}
\eeq
where the normalization of the operator trace is fixed by requiring $\Tr
\hat{\Delta} (x)=1$. Using
(\ref{maptophi}), (\ref{map}) and (\ref{integration}) it is
straightforward to show that the collection of operators $\hat\Delta(x)$ for
$x\in\IR^D$ form an orthonormal set,
\beq
\Tr\Bigl(\hat\Delta(x)\,\hat\Delta(y)\Bigr)=\delta ^{D} (x-y) \ ,
\label{orthoset}\eeq
from which it follows that the inverse of the map (\ref{map}) is given by
\beq
\phi(x)=\Tr\Bigl(\hat\phi\,\hat\Delta(x)\Bigr) \ .
\label{inversemap}\eeq
Therefore, there is a one-to-one correspondence between fields (of sufficiently
rapid decrease at infinity) and operators. This can be thought of as an analog
of the operator-state correspondence of local quantum field theory.

Using these definitions, one can define a scalar quantum field theory on
a noncommutative spacetime as
\beqa
Z &=&
\int \dd \hat{\phi} ~\ee ^{- S\bigl[\hat{\phi}\bigr]} \nonumber\\
S\left[\hat{\phi}\right] &=&\Tr
\left( \frac{1}{2}\,\left[\hat{\del}_\mu\,,\, \hat{\phi}\right]^2
+ \frac{1}{2}\,\hat{\phi} ^2+ \frac{1}{4 \mbox{!}}\,\hat{\phi} ^4 \right) ,
\label{noncomdef}
\eeqa
where the measure
$\dd \hat{\phi} $ is nothing but the ordinary path integral measure
${\cal D} \phi$. The difference between noncommutative quantum field theory
and ordinary field theory comes from the different products of fields that are
used in the two cases. Suppose that $\hat{\phi}_3 = \hat{\phi}_1 \hat{\phi}_2
$, where $\hat{\phi}_i =
\int \dd ^ D x ~ \phi_i (x)\,\hat{\Delta} (x)$. It is straightforward to
compute that the corresponding scalar field $\phi_3(x)$ is given by
\beqa
\phi _3 (x) &=& \Tr\Bigl(\hat\phi_1\hat\phi_2\,\hat\Delta(x)\Bigr)\n&=&
\frac1{\pi ^D|\det\theta |}\,\int\!\!\!\int\dd ^D y ~\dd ^D z ~
\phi _1 (y)\,\phi _2 (z)~
\ee ^{- 2 i (\theta^{-1})_{\mu\nu} (x-y)_\mu (x-z)_\nu } \n
&=& \phi _1 (x)\,\exp\left(\frac i{2} \,
\overleftarrow{\del_\mu} \theta_{\mu\nu}
\overrightarrow{\del_\nu}\right)\,
\phi _2 (x) \n
&\defeq& \phi _1 (x) \star \phi _2 (x) \ ,
\label{starproduct}\eeqa
where in the second line
we have assumed that $\theta_{\mu\nu}$ is an invertible matrix. The
multiplication in (\ref{starproduct}) is called the star or Moyal product of
the fields $\phi_1(x)$ and $\phi_2(x)$. It is associative but not commutative.
For $\theta _{\mu\nu} = 0$, it reduces to the ordinary product of functions and
the theory (\ref{noncomdef}) to the usual
scalar field theory (\ref{scalarft}).

One can easily show that $\int \dd ^D x ~\phi_i (x) \star
\phi_j (x) \star \cdots \star \phi_k (x)$ is invariant
under cyclic permutation of the fields, justifying the representation of the
trace in (\ref{integration}). Since
\beq
\int \dd ^D x ~ \phi_1 (x) \star
\phi_2 (x) =
\int \dd ^D x ~ \phi_1 (x)\,\phi_2 (x) \ ,
\eeq
noncommutative field theory and ordinary field theory are identical at the
level of free fields. The differences come when one considers the interaction
term
\beqa
\Tr\left(\hat{\phi} ^4\right)
&=& \int \dd ^D x ~\phi (x) \star \phi (x) \star \phi (x) \star \phi (x) \n
&=& \int \left(\prod _{i=1}^4 \frac{\dd ^D k _i}{(2\pi)^D} \right) ~
(2\pi)^D\,\delta^D\left(\sum _{i=1}^4 k_i \right)
\left (\prod _{i=1}^4\tilde\phi (k_i) \right)
{}~\ee ^{- \frac{i}{2}\,\theta_{\mu \nu} \sum _{i<j} k_{\mu i} k_{\nu j}} \ .\n
\label{four-int}
\eeqa
We see that the vertex in the noncommutative field theory contains a momentum
dependent phase factor. The interaction is therefore non-local.
Naively one might expect that the effect of the
noncommutativity becomes negligible at energy scales much
smaller than $|\theta_{\mu\nu}| ^{-1/2}$. This is not quite the case.
For example, it has been shown that the noncommutativity
drastically alters the infrared dynamics of the theory \cite{MRS,UVIR}.
The perturbative renormalizability of noncommutative
scalar field theory has been studied in Ref.~\cite{ChepelevRoiban}.

\subsection{Noncommutative Yang-Mills theory}
\label{NCYM_con}

Let us now turn to gauge theory on a noncommutative spacetime.
A hermitian operator corresponding to a
U($p$) gauge field $A_\mu (x)$ can be introduced as
\beq
\hat{A}_\mu = \int \dd ^ D x ~ \hat{\Delta} (x) \otimes A_\mu (x) \ ,
\label{hatAdef}
\eeq
where $\hat{\Delta} (x)$ is given by (\ref{map}).
The action can be written as
\beqa
\label{contaction_op}
S&=&  \Tr\,\tr_{(p)}\left( \left[\hat{\del}_\mu\,,\, \hat{A}_\nu \right]
- \left[\hat{\del}_\nu\,,\, \hat{A}_\mu \right]
+ i \, \left[\hat{A}_\mu\,,\, \hat{A}_\nu \right] \right)^2  \\
&=& \int
\dd ^ D x ~ \tr_{(p)}\Bigl(F_{\mu\nu}(x) \star F_{\mu\nu}(x)\Bigr) \ ,
\label{contaction0}
\eeqa
where
\beq
F_{\mu\nu}= \del _\mu A_\nu - \del _\nu A_\mu
+ i \, (A_\mu \star A_\nu - A_\nu \star A_\mu) \ .
\label{fieldstrengthcont}\eeq
As in (\ref{integration}),
we use the symbol $\Tr$ to denote the operator trace over coordinates, while
$\tr_{(p)}$ denotes the (finite-dimensional) trace in the fundamental
representation of the U$(p)$ gauge group. The action (\ref{contaction0}) is
invariant under the ``star-gauge'' transformation
\beq
A_\mu (x)
\mapsto  g(x) \star A_\mu (x) \star g(x)^\dag
- i \, g(x) \star \del _\mu g(x)^\dag \ ,
\label{con_gaugetr_x}
\eeq
where the gauge function $g(x)$ is a $p \times p$ matrix field satisfying
\beq
g(x) \star g(x)^{\dag}=\id_p \ ,
\eeq
namely it is star-unitary. Introducing a unitary operator corresponding to the
gauge function $g(x)$ through
\beq
\hat{g} = \int \dd ^ D x ~ \hat{\Delta} (x) \otimes g (x) \ ,
\label{hatgdef}\eeq
the gauge transformation (\ref{con_gaugetr_x})
can be written in terms of Weyl operators as
\beq
\hat{A}_\mu \mapsto
\hat{g}\,\hat{A}_\mu\,\hat{g}^{\dag}
- i \, \hat{g} \left[ \hat{\del}_\mu\,,\, \hat{g}^{\dag} \right] \ ,
\label{con_gaugetr_op}
\eeq
under which the action (\ref{contaction_op}) is invariant.

As can be seen from the action (\ref{contaction0}), three-point and four-point
gauge interactions exist even for the simplest case of a U(1) gauge group. It
is known that noncommutative U(1) gauge theory in four dimensions is
asymptotically free, and in fact that its beta-function coincides exactly
with that of large $N$ Yang-Mills theory (See eq.~(\ref{1loopbeta})).
This can be understood as a consequence of Morita equivalence
of noncommutative gauge theories, which we will derive in Section \ref{Morita}.
One form of Morita equivalence states that the non-abelian structure of the
gauge group can be absorbed into the noncommutativity of spacetime. Note in
this regard that Eq.~(\ref{hatAdef})
already has a suggestive form, as it shows that the spacetime indices and the
gauge group indices are treated on the same footing in noncommutative geometry.

\subsection{Star-gauge invariant observables}
\label{observables}

In Ref.~\cite{IIKK}, star-gauge invariant observables in noncommutative
Yang-Mills theory have been discovered using a twisted large $N$ reduced model.
In this subsection, we construct star-gauge invariant observables
in the present general formalism following Ref.~\cite{AMNS}. For this, we first
note that a novel feature of the star-product is that spacetime translations
can be represented on fields via star-conjugation by functions on spacetime.
Let us consider a star-unitary function $S_v(x)$ with the property\footnote{In
principle, $S_v(x)$ can be a $p\times p$ matrix-valued function, but we shall
find below that it is proportional to the unit matrix.}
\beq
S_v(x) \star g(x) \star S_v (x)^\dag = g(x+v)
\label{Svprop}
\eeq
for arbitrary functions $g(x)$,
where $v$ is a real $D$-dimensional vector.
It is straightforward to show that a necessary
and sufficient condition for such a function to exist is
\beq
\theta _{\mu\nu}\,\del _\nu S_v (x) = i\,v_\mu S_v (x) \ .
\label{condSv}
\eeq
In particular, when $\theta _{\mu\nu}$ is invertible,
the solution to (\ref{condSv}) exists for arbitrary
$v_\mu$ and is given by the plane wave
\beq
S_v (x) = \ee ^{i k _\mu  x_\mu}\,\id _p
{}~~~~~~{\rm with}~~k _\mu = (\theta
^{-1})_{\mu\nu}\,v_\nu \ .
\label{Sv}
\eeq
The uniqueness of the function (\ref{Sv}) is immediate. If $S_v'(x)$ is another
function with the property (\ref{Svprop}), then the function
$S_v^{\prime \dag}(x)\star S_v(x)$ star-commutes with all functions $g(x)$. In
particular, by taking $g(x)=\ee^{iw_\mu x_\mu}$
for arbitrary $w _\mu \in\IR$ we may conclude that the function $S_v^{\prime
\dag}(x)\star S_v(x)$ is independent of
$x$, and thus the two functions $S_v(x)$ and $S_v'(x)$ coincide up to some
irrelevant phase factor.

We now construct an analog of the parallel transport operator as
\beqa
{\cal U}(x;C)&=&{\rm P}\,
\exp _\star \left({i\int\limits_C
\dd \xi ^\mu ~A_\mu (x + \xi) }\right)\nonumber\\&\defeq&
1+\sum _{n=1} ^{\infty}i^n \,
\int\limits_0 ^1 \dd \sigma _1 ~
\int\limits_{\sigma_1} ^1 \dd \sigma _2 ~\cdots
\int\limits_{\sigma_{n-1}} ^1 \dd \sigma _n ~
\xi ^{ ' \mu _1}  (\sigma _1)
\cdots
\xi ^{ ' \mu _n} (\sigma _n) \n
& &\times\,A_{\mu _1}\Bigl(x + \xi (\sigma _1)\Bigr)
\star \cdots  \star
A_{\mu _n}\Bigl(x + \xi (\sigma _n)\Bigr)  \ ,
\label{parallel}
\eeqa
where $C$ is an oriented curve in $D$-dimensional spacetime
parametrized by the functions $\xi_\mu (\sigma)$ with $0 \le \sigma \le 1$.
We fix the starting point of the curve to be the origin,
$\xi _\mu (0) = 0$, and denote its endpoint by
$v_\mu = \xi _\mu (1)$ in what follows.
The operator ${\cal U}(x; C)$ can be regarded as a
$p \times p$ star-unitary matrix field depending on $C$.
Under the star-gauge transformation (\ref{con_gaugetr_x}), it transforms as
\beq
{\cal U}(x; C)\mapsto
 g(x)\star {\cal U}(x; C)\star g(x+v )^\dag \ .
\label{gaugecovariant}
\eeq
Star-gauge invariant observables can be constructed
out of (\ref{parallel}) by using the function (\ref{Sv}),
with the property (\ref{Svprop}), as
\beq
{\cal O} (C) =\int \dd ^D x ~  \tr_{(p)} \Bigl( {\cal U}(x; C)
\star S_v(x) \Bigr) \ .
\label{defcalO}
\eeq
In Eq.~(\ref{defcalO}), the parameter $k _\mu$ in
(\ref{Sv}) can be interpreted as the total momentum of the contour $C$.
Note that in the commutative case $\theta_{\mu\nu} = 0$,
gauge invariance requires that the loop be closed irrespectively of the total
momentum. In the noncommutative case, on the other hand,
star-gauge invariance does not require
that the loop be closed, but it does require the separation vector $v_\mu$ of
the loop to be proportional to the total momentum, as is seen from (\ref{Sv}).
The larger the total momentum is, the longer becomes the open loop.
We will study further the dynamics of noncommutative Wilson loops
in Section \ref{properties_hopping} by introducing matter fields
as probes.

Let us now rewrite the star-gauge invariant quantity
(\ref{defcalO}) in terms of Weyl operators.
After a little algebra, the parallel transport operator can be rewritten as
\beq
{\cal U}(x; C)
= \Tr\l  \hat{U}(C) \hat{D}(C)^\dag \,\hat{\Delta}(x)\r \ ,
\label{cU(C)}
\eeq
where we have introduced the unitary operators
\beqa
\hat{U}(C)& =& {\rm P}\,
\exp\left( {\int\limits_C  \dd\xi^\mu~\left( \hat{\del} _\mu +
i \, \hat{A}_\mu\right) }\right) \ , \n
\hat{D}(C) &= & {\rm P}\, \exp \left({\int\limits_C  \dd\xi^\mu~
\hat{\del} _\mu }\right) \ .
\label{cmproduct}
\eeqa
Under a gauge transformation (\ref{con_gaugetr_op}),
$\hat{U}(C)$ transforms as
\beq
\hat{U}(C) \mapsto\hat{g}\,\hat{U}(C)\,\hat{g} ^\dag \ .
\eeq
Note that $\hat{D}(C)$ is nothing but a translation operator
(up to an irrelevant phase factor),
\beq
\hat{D}(C)\,\hat{\Delta}(x)\,
\hat{D} (C)^\dag = \hat{\Delta} (x+v) \ .
\label{Dshift}
\eeq
We also introduce the unitary operator
\beq
\hat{S}_v = \int \dd ^ D x ~ \hat{\Delta} (x)\otimes S_v (x)
=\ee ^ {i k_\mu \hat{x}_\mu}\otimes\id_p
\label{hatSv}\eeq
which satisfies
\beq
\hat{S}_v\,\hat{\Delta} (x)\,\hat{S}_v ^\dag
= \hat{\Delta} (x+v)\otimes\id_p
\label{Sshift}
\eeq
corresponding to (\ref{Svprop}). The Weyl operator description of star-gauge
invariant observables is then given by
\beq
{\cal O} (C) =
 \Tr\, \tr_{(p)} \l \hat{U}(C) \hat{D} (C)^\dag\,\hat{S}_v \r \ .
\label{opendef}
\eeq
Its gauge invariance can be checked directly by noting that the operator
$\hat{D} (C)^\dag\,\hat{S}_v$
in (\ref{opendef}) commutes with $\hat{\Delta} (x)$
due to (\ref{Dshift}) and (\ref{Sshift}).

\subsection{The noncommutative torus}
\label{nctorus}

We now briefly discuss the case where the spacetime is
a $D$-dimensional torus $\IT^D$ instead of $\IR ^D$.
Let us first consider the case discussed in subsection \ref{scalar} and impose
periodic boundary conditions on the scalar field $\phi (x)$ (Twisted boundary
conditions will be studied in the next Section),
\beq
\phi(x+ \Sigma_{\mu a}\,\hat{\mu}) = \phi(x)  \ ,
{}~~~~~~a = 1, \dots , D  \ ,
\label{sigmacompactification_cont}
\eeq
where $\hat\mu$ is a unit vector in the $\mu$-th direction of spacetime, and
$\Sigma_{\mu a}$ is the $D\times D$ period matrix of the torus which is a
vielbein for the metric of $\IT^D$. Here and in the following we use Greek
letters  for spacetime indices and Latin letters for frame indices. Due to the
periodicity (\ref{sigmacompactification_cont}), the momenta $k_\mu$ in the
Fourier mode expansion (\ref{Fourier}) are quantized as
\beq
k _\mu = 2 \pi (\Sigma ^{-1})_{a\mu} m _a \ ,
{}~~~~~~m _ a \in \IZ \ .
\label{momconstr_cont}
\eeq

We define a mapping of fields into operators
as in (\ref{maptophi}) but now with the $\hat{\Delta} (x)$ given by
\beq
\hat{\Delta} (x) =\frac1{|\det  \Sigma|}\,\sum_{\vec{m}\in\IZs^D}
\left(\prod_{ a=1}^D\left(\hat Z _ a\right)^{m_ a}\right)
{}~\ee ^{- \pi i\sum_{ a<b}\Theta_{ ab}
m_ a m_b}~
\ee ^{- 2 \pi i (\Sigma ^{-1})_{ a\mu} m_ a x_\mu } \ ,
\label{map_cont_torus}
\eeq
where the operators
\beq
\hat Z_ a =
\ee ^{2 \pi i (\Sigma ^{-1}) _{ a\mu} \hat{x}_\mu }
\label{Z_cont}
\eeq
satisfy the commutation relations
\beqa
\hat Z_ a\hat Z_b
&=&\ee^{- 2\pi i\Theta_{ ab}}\,\hat Z_b \hat Z_ a \n
\left[ \hat{\del}_\mu\,,\, \hat{Z}_ a \right]
&=& 2 \pi i\,\left(\Sigma ^{-1}\right) _{ a\mu} \,\hat{Z}_ a \ ,
\label{delZcomms}\eeqa
with
\beq
\Theta_{ ab} =
2\pi\,\left(\Sigma ^{-1}\right)_{ a\mu}\,\theta_{\mu\nu}\,
\left(\Sigma ^{-1}\right)_{b\nu}
\label{Thetadimless}\eeq
the dimensionless noncommutativity parameter.
The basis (\ref{map_cont_torus}) has the requisite properties
\beqa
\hat{\Delta} ( x  + \Sigma _{\mu a}\,\hat{\mu})
&=& \hat{\Delta} (x)  \ , ~~~~~  a = 1, \dots , D  \ , \\
\left[\hat \del_\mu\, , \,\hat\Delta (x)\right]
&=& -\partial _\mu\hat\Delta (x) \ .
\label{Deltaprimetransl}\eeqa
Note that the torus $\IT^D$ has a discrete geometrical symmetry
given by its SL$(D,\IZ)$ automorphism group,
under which the period matrix transforms as
$\Sigma \mapsto \Sigma\,\Lambda $ with $\Lambda \in {\rm SL}(D,\IZ)$.
The dimensionless noncommutativity parameter (\ref{Thetadimless}) transforms
as $\Theta \mapsto \Lambda ^{-1}\,\Theta\,(\Lambda ^{-1})^\top$
and the coordinate operators (\ref{Z_cont}) as
$\hat Z_ a \mapsto \prod_{b=1} ^D (\hat  Z_b) ^{(\Lambda ^{-1})_{ a b}}$.
All formulae such as (\ref{map_cont_torus})--(\ref{Thetadimless})
are manifestly SL$(D,\IZ)$ covariant. This symmetry, which persists in the case
of generic twisted boundary conditions on the fields, will be exploited in the
next Section.

Let us now consider star-gauge invariant observables in noncommutative
Yang-Mills theory on the torus $\IT^D$ \cite{AMNS}.
We can define the parallel transport operator
${\cal U}  (x;C) $ as in (\ref{parallel}).
The star-gauge transformation of ${\cal U}  (x;C) $ is given by
(\ref{gaugecovariant}), but now the star-unitary function $g(x)$ is a
single-valued function
on the torus. We can define a star-gauge invariant observable as in
(\ref{defcalO}), where $S_v  (x)$ is a function on the $D$-dimensional torus
which satisfies (\ref{Svprop}) for arbitrary $g(x)$ that we are considering.
One finds that a necessary and sufficient condition is
\beq
S_v\Bigl( x   +2 \pi(\Sigma ^{-1}\,\theta)_{ a\mu}\,\hat{\mu}\Bigr)
=\ee ^{ 2 \pi i (\Sigma ^{-1})_{ a\mu} v_\mu }\,S_v  (x)  \ .
\label{Svequ}
\eeq
The solution can again be given by the plane wave $S_v (x ) = \ee ^{i k_\mu
x_\mu }\,\id_p$, but in the present case $k_\mu$ is quantized as in
(\ref{momconstr_cont}). The condition that must be met is
\beq
v_\mu =  \theta  _{\mu \nu} k _\nu  + \Sigma _{\mu a} n_ a
\label{vk_relation}
\eeq
for some integer-valued vector $n_ a$. Thus we obtain an analog of the Polyakov
line in noncommutative Yang-Mills theory represented by the existence of the
case with non-zero ``winding number'' $n_ a$.

\section{Morita equivalence}
\label{Morita}
\setcounter{equation}{0}

In noncommutative geometry, there is a remarkable geometric equivalence
relation on certain classes of noncommutative spaces known as Morita
equivalence \cite{NCgeom}. Roughly speaking, two spaces are Morita equivalent
if one space can be regarded as a twisted operator bundle over the other space
of a certain topological charge whose fibers are operator algebras. Many
topological quantities are preserved by the Morita equivalence relation. In
particular, K-theory groups are invariant under it, so that two Morita
equivalent spaces
should have a canonical mapping between gauge bundles defined over them. In
noncommutative Yang-Mills theory, this implies a remarkable duality between
gauge theories over different noncommutative tori \cite{Morita}, for example,
which relates a Yang-Mills theory with background magnetic flux to a gauge
theory with gauge group of lower rank and no background flux. It allows one to
interpolate continuously, through noncommutative Yang-Mills theories, between
two ordinary Yang-Mills theories with gauge groups of different rank and
appropriate
background magnetic fluxes. Furthermore, in certain instances, there is the
remarkable fact that the non-abelian nature of a gauge group can be absorbed
into the noncommutativity of spacetime by mapping a U$(p)$ gauge theory with
multi-valued gauge fields to a U(1) gauge theory with single-valued fields on a
dual noncommutative torus. In string theory, Morita equivalence coincides with
the phenomenon of T-duality of open string backgrounds in the presence of
D-branes \cite{SW,Morita}.

The nonperturbative formulation of generic noncommutative Yang-Mills theories
that we shall present in Section~\ref{lattice} relies crucially on the Morita
equivalence property. We shall therefore present a detailed and rather general
derivation of the Morita equivalence relation for noncommutative tori from the
point of view adopted in the previous Section to noncommutative Yang-Mills
theory. We shall see that, in the present formalism, the equivalence has a
natural geometrical interpretation as a change of basis for the map between
fields and Weyl operators for the noncommutative geometry. We will also show
that this duality holds at the level of star-gauge invariant
observables in the two gauge theories, thereby providing a complete equivalence
of the noncommutative quantum field theories.

\subsection{Twisted gauge theory on the noncommutative torus}
\label{GTderivation1}

Consider a U($p$) gauge theory on a noncommutative torus $\IT_\Theta^D$ with
gauge fields of non-vanishing topological charge.
Such gauge fields are not single-valued functions on the
torus. Instead, they must be regarded as functions on the universal covering
space $\IR^D$ with the twisted boundary conditions
\beq
A_\mu (x+ \Sigma _{\nu a}\,\hat{\nu}) =
\Omega _ a(x) \star  A_\mu (x) \star\Omega _ a(x)^\dag
- i \, \Omega _ a(x) \star \del_\mu \Omega _ a(x)^\dag \ .
\label{con_bc_gen_higher}
\eeq
The transition functions $\Omega_ a (x)$
are star-unitary $p\times p$ matrices, which we decompose into an abelian part
and an SU$(p)$ part via the gauge choice
\beq
\Omega_ a (x)=
\ee^{i\alpha_{ a\mu} x_\mu}\otimes\Gamma_ a \ ,
\label{con_omega_explicit_higher}
\eeq
where $\alpha$ is a real-valued constant $D\times D$ matrix
satisfying $(\alpha \Sigma ) ^\top = - \alpha \Sigma$,
and $\Gamma_ a$ are SU$(p)$ matrices. The matrix $\alpha$
will account for the abelian fluxes of the gauge fields. The
antisymmetry of the matrix $\alpha \Sigma$
implies that the transition function $\Omega_ a(x)$ has periodicity
$\Omega _ a (x + \Sigma _{\mu  a}\,\hat{\mu})= \Omega _ a (x)$.

Consistency of the conditions (\ref{con_bc_gen_higher}) requires the
$\Omega_ a$ to obey the cocycle condition
\beq
\Omega _ a
(x + \Sigma _{\mu b}\,\hat{\mu}) \star \Omega _ b (x)
= {\cal Z}_{ a b}~
\Omega _ b (x+ \Sigma _{\mu a}\,\hat{\mu}) \star
\Omega _ a (x) \,,
\label{cocycleD_higher}
\eeq
where ${\cal Z}_{ a b}
=\ee^{2\pi i\gamma_{ a b}/p}$ are elements of the
center of the SU($p$) part of the gauge group, with $\gamma$ an
antisymmetric integral $D\times D$ matrix.
In commutative SU($p$) gauge theory, a phase factor such as ${\cal Z}_{ a b}$
would induce a non-abelian 't~Hooft flux for the gauge fields \cite{tHooft}.
In that case, the matrix $\alpha$ in (\ref{con_omega_explicit_higher}) should
be set to zero, and therefore having a non-trivial
${\cal Z}_{ a b}$ is the only way to twist the boundary conditions on the gauge
fields. In commutative U($p$) gauge theory, keeping
both $\gamma$ and $\alpha$ non-zero is redundant because one can eliminate the
U(1) twists $\alpha$ by $\IZ_p$-valued phase factors according to the global
decomposition ${\rm U}(p)={\rm U}(1)\times{\rm SU}(p)/\IZ_p$,
and one can set either $\gamma= 0$ or $\alpha = 0$
without loss of generality. However, this is not quite the case
in noncommutative U($p$) gauge theory, as we will see.

Eq.~(\ref{cocycleD_higher}) implies that the matrices $\Gamma_ a$
mutually commute up to some phases,
\beq
\Gamma_ a \Gamma_ b
=\ee^{2\pi iQ_{ a b}/p}\,\Gamma_ b \Gamma_ a \ .
\label{GammaalgD_higher}
\eeq
Taking the determinant of both sides of (\ref{GammaalgD_higher})
shows that the antisymmetric matrix $Q$ has elements $Q_{ a b}\in\IZ$.
Furthermore, eq. (\ref{cocycleD_higher}) gives the consistency condition
\beq
Q
=\frac p{2\pi}\,\Bigl(2\alpha\,\Sigma - \alpha\,\theta\,\alpha ^\top \Bigr)
- \gamma  \ ,
\label{Qalpharel_higher}
\eeq
where here and in the following we use a matrix multiplication convention.
The integer $Q_{ a b}$ is the magnetic flux of the gauge field through the
surface formed by the $ a$-th and $ b$-th cycles of the torus. The central
extension of the cocycle relation (\ref{cocycleD_higher}) shifts $Q$ by the
matrix $\gamma$ of non-abelian 't~Hooft fluxes representing the associated
principal curvatures.

The fact that the trivial configuration $A_\mu (x)=0$ is not a solution of
(\ref{con_bc_gen_higher}) motivates the introduction of a fixed, multi-valued
background abelian gauge field $A^{(0)}_\mu (x)$ defined by
\beq
A^{(0)}_\mu (x) =\frac12\,F_{\mu\nu}\,x_\nu\otimes\id_p \ ,
\label{con_u_0_higher}
\eeq
where $F$ is a real-valued constant antisymmetric $D\times D$ matrix.
The twisted boundary conditions (\ref{con_bc_gen_higher}) for the gauge field
$A^{(0)}_\mu (x)$ are equivalent to the relations
\beqa
\alpha&=& - \Sigma ^ \top F\,\frac1{2\id_D+\theta F}  \n
F&=&2\alpha ^\top \,\frac1{\Sigma -\theta\alpha ^\top } \ .
\label{alphaFrels_higher}
\eeqa
We then decompose $A_\mu (x)$ into a part representing a particular solution to
the twisted boundary conditions (\ref{con_bc_gen_higher}) and a part
representing the fluctuations around the fixed background as
\beq
A_\mu (x)=A^{(0)}_\mu (x) +  {\cal A}_\mu (x) \ ,
\label{decomposeA_field}
\eeq
where ${\cal A}_\mu (x)$ satisfies the constraints
\beq
{\cal A}_\mu (x+ \Sigma _{\nu a}\,\hat{\nu}) =
\Omega _ a (x) \star {\cal A}_\mu (x) \star
\Omega _ a (x) ^\dag \ .
\label{con_bc2_higher}
\eeq
This means that the field ${\cal A}_\mu(x)$ is an adjoint section of the
corresponding gauge bundle over the noncommutative torus.

Let us now consider the noncommutative Yang-Mills action
\beq
S=\frac1{g^2}\,\int
\dd ^ D x ~ \tr_{(p)}\Bigl(F_{\mu\nu}(x)  - f_{\mu\nu}\,\id_p\Bigr)_\star^2 \ ,
\label{contaction_higher}
\eeq
where the noncommutative field strength tensor $F_{\mu\nu}(x)$ is defined in
(\ref{fieldstrengthcont}). The constant background tensor field $f_{\mu\nu}$
will be specified later, and the integration is taken over the torus $\IT^D$.
Using the decomposition (\ref{decomposeA_field}),
the action (\ref{contaction_higher}) can be written as
\beq
S=\frac1{g^2}\,\int \dd ^ D x ~ \tr_{(p)}\Bigl({\cal F}_{\mu\nu}(x)
+ F_{\mu\nu}^{(0)}- f_{\mu\nu}\,\id_p\Bigr)_\star^2 \ ,
\label{contaction2_higher}
\eeq
where
\beqa
{\cal F}_{\mu\nu}&=&
 \del _\mu {\cal A} _\nu -\del_\nu{\cal A}_\mu+
i \,\left( {\cal A} _\mu\star {\cal A}
_\nu-{\cal A} _\nu\star {\cal A} _\mu\right) \n
&~& +\,i \,\left( A ^{(0)}_\mu\star {\cal A} _\nu
- {\cal A}_\nu\star A ^{(0)}_\mu\right)
-i\,\left(A^{(0)}_\nu\star{\cal A}_\mu-{\cal A}_\mu\star A^{(0)}_\nu\right)
\label{calFdef_higher1}\\
F_{\mu\nu}^{(0)}&=&  \del _\mu  A ^{(0)}_\nu-\del _\nu  A ^{(0)}_\mu
+ i \,\left( A ^{(0)}_\mu\star A ^{(0)}_\nu-A ^{(0)}_\nu\star A ^{(0)}_\mu
\right)=\left(F+\frac14\,F\theta F\right)_{\mu\nu}\otimes\id_p \ .\n
\label{calFdef_higher2}
\eeqa
{}From (\ref{alphaFrels_higher}) and the identity
\beq
\left(\frac1{\id_D-\theta\alpha ^\top \Sigma ^{-1}}\right)^2\otimes\id_p
=\left(\id_D+ \frac1{2} \, \theta F\right)^2\otimes\id_p=
\id_D\otimes\id_p+\theta F^{(0)}
\label{alphathetaid}\eeq
it follows that the relation (\ref{Qalpharel_higher}) is equivalent to
\beq
\Sigma ^\top\,F^{(0)}\,\Sigma
=2 \pi \, \frac1{p\id_D+(Q+\gamma)\Theta}\,\Bigl(Q+\gamma\Bigr)
\otimes\id_p \ .
\label{calFQrel}\eeq
Eq.~(\ref{calFQrel}) gives the relationship between the central curvatures, the
topological charges, and the 't~Hooft fluxes of the gauge field configurations.
Requiring that ${\cal A} _\mu (x)= 0$ be the vacuum field configuration of the
theory up to a gauge transformation fixes $f_{\mu\nu}\,\id_p=F_{\mu\nu}^{(0)}$,
so that the action becomes
\beq
S=\frac1{g^2}\,\int \dd ^ D x ~ \tr_{(p)}\Bigl({\cal F}_{\mu\nu}(x)\star{\cal
F}_{\mu\nu}(x)\Bigr) \ .
\label{contaction3_higher}
\eeq

\subsection{Transformation of the action}
\label{GTderivation2}

In this subsection, we will derive the Morita equivalence relation between
noncommutative Yang-Mills theories in arbitrary even dimension $D=2d$.
Specifically, we will show that the theory (\ref{contaction_higher}) with the
constraint (\ref{con_bc_gen_higher}) is equivalent to another
noncommutative gauge theory on a torus with single-valued gauge fields.
For this, we first rewrite the gauge theory of the previous subsection in terms
of Weyl operators. We introduce the hermitian operator $\hat{A} _\mu $ as in
(\ref{hatAdef}), and define the constant abelian background $\hat{A} _\mu
^{(0)} $ and the transition function $\hat{\Omega} _ a$ similarly.
The explicit gauge choice (\ref{con_omega_explicit_higher})
corresponds to the unitary operator
\beq
\hat{\Omega}_ a =\ee^{i\alpha_{ a\mu}\hat{x}_\mu}\otimes\Gamma_ a \ ,
\label{con_omega_explicit_higher2}
\eeq
and, corresponding to (\ref{decomposeA_field}), we decompose
$\hat{A} _\mu $ as
\beq
\hat{A}_\mu = \hat{A} ^{(0)} _\mu +  \hat{\cal A}_\mu \ .
\label{decomposeA_op}
\eeq
The action
(\ref{contaction_higher}) or (\ref{contaction3_higher})
can be written as
\beqa
\label{actionA_op}
S&=&\frac{1}{g^2} ~
\Tr\, \tr_{(p)} \left( \left[\hat{\del}_\mu\,,\, \hat{A}_\nu \right]
- \left[\hat{\del}_\nu\,,\, \hat{A}_\mu \right]+i \,
 \left[\hat{A}_\mu\,,\, \hat{A}_\nu \right] - f_{\mu\nu}\,\id_p\right)^2  \\
&=& \frac{1}{g^2}~
\Tr\, \tr_{(p)} \left(
\left[ \hat{\nabla} ^{(0)} _\mu\,,\, \hat{\cal A}_\nu \right]
- \left[ \hat{\nabla} ^{(0)} _\nu\,,\, \hat{\cal A}_\mu \right]
+ i \, \left[ \hat{\cal A}_\mu\,,\, \hat{\cal A}_\nu \right]
\right)^2  ,
\label{actionAhat_op}
\eeqa
where $f _{\mu\nu}$ may be given in terms of $\hat {A} _\mu ^{(0)}$ as
\beq
f _{\mu\nu}\,\id_p=
\left[\hat{\del}_\mu\,,\, \hat{A}_\nu ^{(0)} \right]
- \left[\hat{\del}_\nu\,,\, \hat{A}_\mu ^{(0)} \right]
+ i \, \left[\hat{A}_\mu ^{(0)}\,,\, \hat{A}_\nu ^{(0)}\right]
\eeq
and we have introduced the fiducial constant curvature connection
\beq
\hat{\nabla}^{(0)}_\mu  = \hat{\del} _\mu + i \, \hat{A} ^{(0)} _\mu \ .
\label{defnabla}
\eeq
The constraint (\ref{con_bc_gen_higher}) and the equivalent one
(\ref{con_bc2_higher}) can be written in terms of operators as
\beqa
\label{constraintA_op}
\ee ^{ \Sigma _{\nu  a} \hat{\del} _\nu}\,\hat{A} _\mu~
\ee ^{-  \Sigma_{\nu  a} \hat{\del} _\nu}
&=& \hat{\Omega} _ a\,\hat{A} _\mu\,\hat{\Omega} _ a ^ \dag
- i \,  \hat{\Omega} _ a
\left[ \hat{\del} _\mu\,,\, \hat{\Omega} _ a ^ \dag \right] \\
\ee ^{  \Sigma _{\nu  a} \hat{\del} _\nu}\,\hat{\cal A} _\mu~
\ee ^{-  \Sigma _{\nu  a} \hat{\del} _\nu}
&=& \hat{\Omega} _ a\,\hat{\cal A} _\mu\,\hat{\Omega} _ a ^ \dag \ .
\label{constraintAhat_op}
\eeqa
Our next task is to solve this constraint for the gauge field configurations.

First of all, we see from (\ref{GammaalgD_higher})
that the $\Gamma_ a$ are twist eating solutions
for SU$(p)$. For generic rank $p$ and flux matrix $Q$,
these matrices may be constructed as follows \cite{twisteater}.
For this, it is convenient to use the discrete SL$(D,\IZ)$ symmetry of $\IT^D$
to transform the twist eaters into matrices $\Gamma _ b ^\circ$ through
\beq
\Gamma _ a  = \prod _{ b=1}^D (\Gamma _ b ^\circ )^{\Lambda_{ b a}} \ ,
\label{new_old_Gamma}
\eeq
where $\Lambda \in {\rm SL}(D,\IZ)$. The transformed twist eaters $\Gamma _ a
^\circ$ satisfy the commutation relations
\beq
\Gamma_ a ^\circ \Gamma_ b ^\circ
=\ee^{2\pi iQ_{ a b}^\circ/p}\,
\Gamma_ b ^\circ \Gamma_ a ^\circ \ ,
\label{newtwistcom}
\eeq
with $Q = \Lambda ^\top Q^\circ \Lambda$.
We can now choose $\Lambda$ so that the matrix $Q^\circ$ takes a canonical
skew-diagonal form
\beq
Q^\circ
=\pmatrix{0&-q_1& & & \cr q_1&0& & & \cr & &\ddots& & \cr & & &0&-q_d\cr & &
&q_d&0\cr} \ .
\label{Qdiag}\eeq
Given the $d$ independent fluxes $q_i\in\IZ$, we introduce the three integers
\beq
p_i={\rm gcd}(q_i,p)~~~~~~,~~~~~~\tilde p_i=\frac p{p_i}
{}~~~~~~,~~~~~~\tilde q_i=\frac{q_i}{p_i} \ .
\label{pitildedefs}\eeq
By construction, $\tilde p _i$ and $\tilde q _i$ are co-prime.
A necessary and sufficient condition for the existence of solutions to
(\ref{newtwistcom}) is \cite{twisteater}
that the integer $\tilde p_1\cdots\tilde p_d$, which is the dimension of the
irreducible representation of the Weyl-'t~Hooft algebra
(\ref{GammaalgD_higher}), divides the rank $p$. This is a condition which must
be met by the geometrical parameters of the given constant curvature bundle. In
that case we write
\beq
p=\tilde p_0\,\prod_{i=1}^d\tilde p_i
\label{tildep0def}\eeq
and the twist eating solutions may then be given on the subgroup ${\rm
SU}(\tilde p_1)\otimes\cdots\otimes{\rm SU}(\tilde p_d)\otimes{\rm SU}(\tilde
p_0)$ of SU$(p)$ as
\beqa
\Gamma_{2i-1}^\circ&=&\id_{\tilde p_1}\otimes\cdots\otimes V_{\tilde
p_i}\otimes\cdots\otimes\id_{\tilde p_d}\otimes\id_{\tilde
p_0}\n\Gamma_{2i} ^\circ
&=&\id_{\tilde p_1}\otimes\cdots\otimes\left(W_{\tilde
p_i}\right)^{\tilde q_i}\otimes\cdots\otimes\id_{\tilde p_d}\otimes\id_{\tilde
p_0}
\label{twisteat}\eeqa
for $i=1,\dots,d$. Here $V_p$ and $W_p$ are the SU$(p)$ shift and clock
matrices
\beq
V_p=\pmatrix{0&1& & &0\cr &0&1& & \cr& &\ddots&\ddots& \cr& & &\ddots&1\cr 1& &
& &0\cr}~~~~~~,~~~~~~W_p=\pmatrix{1& & & & \cr &\ee^{2\pi i/p}& & & \cr&
&\ee^{4\pi i/p}& & \cr& & &\ddots& \cr & & & &\ee^{2\pi i(p-1)/p}\cr}
\label{clockshift}
\eeq
obeying $V_pW_p=\ee^{2\pi i/p}\,W_pV_p$.
We can then obtain twist eaters $\Gamma _ a$ satisfying
(\ref{GammaalgD_higher}) by using the relation (\ref{new_old_Gamma}).
Note that the space of matrices which commute with the $\Gamma_ a$
is the gl$(\tilde p_0,\IC)$ subspace of ${\rm gl}(p,\IC)$ generated by the
matrices $\id_{\tilde p_1}\otimes\cdots\otimes\id_{\tilde p_d}\otimes Z_0$,
$Z_0 \in{\rm gl}(\tilde p_0 , \IC)$.

Since the set $\{(V_p)^{j} (W_p)^{j'}~|~
j,j' \in\IZ_{p}\}$ spans the linear space gl($p , \IC$),
the operator $\hat{\cal A}_\mu$ may be expanded as
\beq
\hat{\cal A} _\mu =
\sum_{\vec{j}} \int \dd^Dk~\ee ^ {i k_\nu\hat{x}_\nu}
\otimes \prod_{ a=1} ^{D}
(\Gamma_ a)^{j_ a} \otimes a_\mu(k , \vec{j}\,) \ ,
\label{hatAexp}\eeq
where the $\tilde{p}_0 \times \tilde{p}_0 $ matrix-valued coefficients
$a_\mu(k , \vec{j}\,)$ are periodic functions of $\vec{j}$ with the periodicity
$j_ a \sim j_ a + \tilde P _{ a b} $, $ b = 1,\dots , D$. The matrix $\tilde P$
will be specified below. The constraint (\ref{constraintAhat_op}) then implies
that $a_\mu(k , \vec{j}\,)$ vanishes unless
\beq
\zeta_ a + \frac1{p} \, Q_{ a b} j_ b =  n_ a  \in \IZ \ ,
\label{amuconstr}
\eeq
where we have defined the $D$-dimensional real-valued vector
\beq
\zeta_ a =  \frac1{2\pi} \,
k_\nu\Bigl(\Sigma + \theta \alpha ^\top\Bigr)_{\nu  a}  \ .
\label{Xtok}
\eeq
We will now find the general solution ($\zeta _ a$, $j_ b$)
to eq.~(\ref{amuconstr}). Introducing the $D\times D$ integral matrices
\beq
\tilde P ^\circ
=\pmatrix{\tilde p_1&0& & & \cr0&\tilde p_1& & & \cr
& &\ddots& & \cr & & &\tilde p_d&0\cr & &
&0&\tilde p_d\cr}~~~~~~,~~~~~~\tilde Q  ^\circ
=\pmatrix{0&-\tilde q_1& & & \cr \tilde q _1&0& & &
\cr & &\ddots& &
\cr & & &0&-\tilde q_d\cr & & &\tilde q_d&0\cr}  \ ,
\eeq
and $\tilde{P} = \Lambda ^{-1} \tilde P ^\circ \Lambda '$,
$\tilde{Q} = \Lambda ^{\top} \tilde Q ^\circ \Lambda '$,
we can write
\beq
Q=p\,\tilde Q \tilde P ^{-1} \ ,
\label{QtildeQ}\eeq
where $\Lambda ' \in {\rm SL}(D,\IZ)$ will represent
the automorphism symmetry group of the resulting theory that we shall find.
Then eq.~(\ref{amuconstr}) can be written as
$\zeta _ a = m_ c (\tilde P ^{-1})_{ c a}$,
where $m_ c$ is a $D$-dimensional integral vector which satisfies
\beq
m_ c = n_ b\tilde P_{ b c} + j _ b \tilde Q _{ b c}  \ .
\label{m_constraint}
\eeq

We will next show that for any given integer $m_ c$, there exists a set of
integers ($n_ b$, $j_ b$) which satisfies (\ref{m_constraint}).
For this, we note that since $\tilde p_i$ and $\tilde q_i$ are relatively
prime, there exists a set of integers $(a_i,b_i)$ such that
\beq
a_i \tilde p_i+ b_i \tilde q_i =1~~~~~~,~~~~~~i=1,\dots,d \ .
\label{apbqpi}
\eeq
Introducing the $D\times D$ integral matrices
\beq
A^\circ
=\pmatrix{a_1&0& & & \cr0&a_1& & & \cr & &\ddots& & \cr & & &a_d&0\cr & &
&0&a_d\cr}~~~~~~,~~~~~~
B^\circ
=\pmatrix{0&-b_1& & & \cr b_1&0& & & \cr & &\ddots& &
\cr & & &0&-b_d\cr & & &b_d&0\cr}  \ ,
\eeq
and
$A = (\Lambda ')^{-1} A ^\circ \Lambda $,
$B = (\Lambda ')^{-1} B ^\circ (\Lambda ^{-1})^\top$, we have
\beq
A\tilde P+B\tilde Q=\id_D \ .
\label{ABtilderel}\eeq
The integers $n_ b = m _ a A_{ a b}$
and $j_ b = m _ a B_{ a b}$ then give a solution to (\ref{m_constraint}).
It is easy to see that for any given $m_ a$,
the $j_ b$ satisfying (\ref{m_constraint}) are unique
up to their periodicity. Thus, we find that
the general solution to eq.~(\ref{amuconstr}) can be given by
\beq
\zeta _ a  = m_ b\left(\tilde P ^{-1}\right)_{ b a} \ ,  ~~~~
j_ a  = m_ b B_{ b a} \ ~~~~~~~\forall m_ b \in \IZ \ .
\eeq
Solving (\ref{Xtok}) for the momenta we obtain $k_\mu = 2 \pi  m_ a\beta_{ a
\mu}$, where we have defined the $D\times D$ matrix
\beq
\beta=\frac1{(\Sigma + \theta\alpha ^\top)\tilde P} \ .
\label{beta}
\eeq

Therefore, the most general gauge field configuration satisfying
the constraint (\ref{constraintAhat_op}) is given as
\beq
\hat{\cal A}_\mu  =\sum_{\vec m \in \IZs ^D}
\ee^{\pi i\sum_{ a< b}\Theta'_{ a b}m_ a m_ b}
\left(\prod_{ a=1}^D\left(\hat Z_ a '\right)^{m_ a}\right)
\otimes  \tilde a_\mu (\vec m)  \ ,
\label{AiZD}
\eeq
where $\tilde a_\mu(\vec m)$ are $\tilde p_0 \times \tilde p_0$
matrix-valued coefficients and
\beq
\hat Z_ a '=\ee^{2\pi i\beta_{ a\mu}\hat{x}_\mu}
\otimes\prod_{ b=1}^D(\Gamma_ b)^{B_{ a b}} \ .
\label{Ziansatz}\eeq
Hermiticity of $\hat{\cal A}_\mu $ requires
$\tilde a_\mu (- \vec  m) = \tilde a_\mu (\vec m)^\dag$.
The operators (\ref{Ziansatz}) obey the commutation relations
\beqa
\hat Z_ a '\hat Z_ b '
&=&\ee^{- 2\pi i\Theta_{ a b}'}\,\hat Z_ b'\hat Z_ a '  \ ,\n
\Bigl[\hat{\nabla}_\mu^{(0)}\,,\,\hat Z_ a ' \Bigr]&=& 2\pi i
\,\left(\Sigma ^{' -1}\right)_{ a\mu}\,\hat Z_ a '  \ ,
\label{nablaZcomm}\eeqa
where
\beqa
\Theta' & = &
\frac1{\Theta\left(\tilde Q+\frac1p\,\gamma \tilde P \right)-\tilde P}
\,
\left[\left(A-\frac1p\,B\gamma\right)\Theta+B\right] ^\top  \ ,
\label{thetaprimeD}    \\
\Sigma^{' } &=&  \Sigma  \,
\left[ \Theta \left(\tilde Q+
\frac1p\, \gamma \tilde P\right) -\tilde P \right]  \ .
\label{metrictransf}
\eeqa
The commutation relations (\ref{nablaZcomm}) are the same
as (\ref{delZcomms}) with the replacements
$\hat Z \to \hat Z '$, $\hat \del  \to \hat \nabla  ^{(0)}$,
$\Theta \to \Theta ' $ and $\Sigma \to \Sigma ' $.
We may therefore define a basis $\hat{\Delta} ' (x ')$ for the mapping of
fields into operators as in (\ref{map_cont_torus}).
The $x'_\mu\in\IR$ are interpreted as coordinates on a new,
dual torus with period matrix $\Sigma '$ given by (\ref{metrictransf}).
Furthermore, the dimensionful noncommutativity parameters $\theta_{\mu\nu}'$
which define the star-product are given by the
dimensionless parameters (\ref{thetaprimeD}) as
\beq
\theta'=\frac1{2\pi}\,\Sigma'\,\Theta'\,\Sigma'^\top \ .
\label{thetaprimesmall}\eeq

The expansion (\ref{AiZD}) then becomes
\beq
\hat{\cal A}_\mu = \int \dd ^D x ' ~ \hat{\Delta} ' (x ')\otimes
{\cal A}_\mu ' (x') \ ,
\label{con_newexpansion_higher}
\eeq
where ${\cal A}_\mu '(x ')$ can be regarded as a single-valued U($\tilde{p}_0$)
gauge field on the noncommutative torus $\IT_{\Theta'}^D$ of total volume
$|\det\Sigma'|$. The operator trace $\Tr'$, with $\Tr'\,\hat\Delta'(x')=1$,
over the coordinates of this new torus is related to the original trace $\Tr$
through
\beq
\Tr'\,\tr_{(\tilde
p_0)}=  \frac{\tilde p _0}{p} \,
\left|\frac{\det \Sigma^\prime}{\det \Sigma}\right|
{}~\Tr\,\tr_{(p)} \
{}.
\label{tracechange}\eeq
Using (\ref{actionAhat_op}) and (\ref{tracechange}) we arrive at the canonical
form of the noncommutative Yang-Mills action for the gauge field in
(\ref{con_newexpansion_higher}),
\beq
S=\frac1{g'^2}\,\int\dd^Dx'~\tr_{(\tilde p_0)}\Bigr({\cal
F}_{\mu\nu}'(x')\star'{\cal F}_{\mu\nu}(x')\Bigr) \ ,
\label{newYMaction}\eeq
where
\beq
{\cal F}_{\mu\nu}'=\partial_\mu'{\cal A}_\nu'-\partial_\nu'{\cal A}_\mu'
+ i \,\left({\cal A}_\mu'\star'{\cal A}'_\nu-{\cal A}_\nu'\star'{\cal A}_\mu'
\right) \ ,
\label{calFprime}\eeq
$\star '$ denotes the star product defined using $\theta '$ instead of
$\theta$, and the new Yang-Mills coupling constant
is given by\footnote{The transformation (\ref{couplingchange}) of the
Yang-Mills coupling constant differs from the standard one which is derived
using string theoretical T-duality. The source of this discrepency owes to the
normalization of the trace Tr on the noncommutative torus $\ITs_\Theta^D$. In
string theory, this trace is normalized according to the standard formula in
noncommutative geometry for the rank of the module of sections of the
fundamental bundle associated to the given Chan-Paton gauge bundle over
$\ITs_\Theta^D$ (This module coincides with the Hilbert space of open string
ground states) \cite{SW,Morita}. In the present field theoretical case, which
does not require the specification of any representation of the Weyl operators,
we have chosen the more natural volume normalization $\Tr\,\id=|\det\Sigma|$.}
\beq
g^{'2}= g^2\,\frac{\left|\det\Bigl[\Theta(Q+\gamma)-p\id_D\Bigr]\right|}
{\tilde p_0\,p^{D-1}} \ ,
\label{couplingchange}\eeq
where we have used (\ref{QtildeQ}).

We have therefore shown that the U$(p)$ gauge theory (\ref{contaction_higher})
on a bundle of topological charges $Q_{ a b}$ and 't~Hooft fluxes
$\gamma_{ a b}$ over the noncommutative torus $\IT_\Theta^D$ (i.e. with gauge
fields obeying twisted boundary conditions) is equivalent to the U$(\tilde
p_0)$ gauge theory (\ref{newYMaction}) on a trivial bundle over the
noncommutative torus $\IT_{\Theta'}^D$ (i.e. with gauge fields obeying periodic
boundary conditions) with noncommutativity parameter matrix defined by
(\ref{thetaprimeD}) and the reduced rank $\tilde p_0$ by (\ref{tildep0def}). In
particular, for $\tilde p_0=1$, the internal matrix structure of the gauge
fields are completely absorbed into the operators $\hat Z_ a'$ which are
regarded as the coordinate generators of $\IT_{\Theta'}^D$, thereby allowing
one to reinterpret the original non-abelian gauge theory as a U(1) gauge theory
on the noncommutative torus $\IT_{\Theta'}^D$ (Note that this statement is true
even for $\theta=0$). On the other hand, when $Q=0$ we have $\tilde p_0=p$ and
the rank of the gauge fields is unchanged. Note that from (\ref{thetaprimeD})
and (\ref{metrictransf}) it follows that,
when $\theta = 0$, the resulting dual gauge theory does not depend on
$\gamma$ and $\alpha$ separately, but only on their combination given by $Q$ in
(\ref{Qalpharel_higher}). This means that keeping both $\gamma$ and
$\alpha$ non-zero is in fact redundant in the commutative case.
But for $\theta \neq 0$, this is not quite the case.

We note that eq.~(\ref{ABtilderel}) and the antisymmetry of the matrices $A
B^\top$ and $\tilde Q ^\top  \tilde P$ implies the matrix identity
\beq
\pmatrix{A&B \cr \tilde Q ^\top & - \tilde P ^\top \cr}
\pmatrix{0&\id _D \cr \id _D  & 0 \cr}
 \pmatrix{A&B \cr \tilde Q ^\top & - \tilde P ^\top \cr} ^\top
= \pmatrix{0&\id _D \cr \id _D  & 0 \cr}   \ .
\label{SO2D}\eeq
It follows that, when $\gamma=0$, the map $\Theta\mapsto\Theta'$ in
(\ref{thetaprimeD}) is the usual SO$(D,D;\IZ)$ transformation that relates
Morita equivalent noncommutative tori \cite{Morita}. From the present point of
view, Morita equivalence is therefore regarded as a change of basis
$\hat\Delta(x)\leftrightarrow\hat\Delta'(x')$ for the mapping between Weyl
operators and fields. The choice $\gamma=0$ is required whenever one wants to
represent the operators $\hat Z_ a'$ on the Hilbert space of sections of the
associated fundamental bundle, as in the case of D-brane applications wherein
the open string wavefunctions transform in the fundamental representation of
the Chan-Paton gauge group \cite{SW}. A choice $\gamma\neq0$ is possible for
representations on the Hilbert space of sections of the associated principal
bundle. Notice also that for $d>1$ the calculation presented above accounts for
only a subset of the possible noncommutative gauge theories, since in that
instance, even in the commutative case, generic bundles over the torus do not
always admit constant curvature connections such as (\ref{defnabla}). However,
the derivation above does allow for arbitrary non-abelian 't~Hooft fluxes
$\gamma$.

\subsection{Star-gauge symmetry and transformation of observables}
\label{observables_morita}

We will now demonstrate explicitly how the star-gauge symmetries of two
Morita equivalent Yang-Mills theories, as well as their star-gauge invariant
observables, are related to each other. We begin with the U$(\tilde p_0)$
theory (\ref{actionAhat_op}) in which all fields obey periodic boundary
conditions.
Its gauge symmetry is
\beq
\hat{\cal A} _\mu \mapsto
\hat{g}\,\hat{\cal A} _\mu\,\hat{g} ^\dag
- i \,
\hat{g}\,\left[ \hat{\nabla} _\mu ^{(0)}\,,\, \hat{g} ^\dag\right] \ ,
\label{gaugeinvU1}
\eeq
where $\hat{g}$ may be written as
\beq
\hat{g} = \int \dd ^D x ' ~ \hat{\Delta} ' (x ')\otimes g'(x ') \ .
\label{con_expansion_g}
\eeq
The gauge function $g'(x')$ is a single-valued, $\tilde p_0 \times \tilde p_0$
star($\star '$)-unitary matrix field,
which parametrizes the usual gauge transformation of
the noncommutative Yang-Mills theory (\ref{newYMaction}). To see how this is
interpreted as the gauge symmetry of the U($p$) theory (\ref{actionA_op}), we
use (\ref{defnabla}) to rewrite (\ref{gaugeinvU1}) as
\beq
\hat{A} _\mu \mapsto
\hat{g}\,\hat{A} _\mu\,\hat{g} ^\dag
- i \, \hat{g}\,\left[ \hat{\del} _\mu\,,\, \hat{g} ^\dag\right] \ ,
\label{gaugeinvUp}
\eeq
where $\hat{A} _\mu$ is given by (\ref{decomposeA_op}).
The invariance of the action (\ref{actionA_op})
under (\ref{gaugeinvUp}) follows by construction.
Note that by the definition (\ref{con_expansion_g}), the operator $\hat g$
obeys
\beq
\ee^{\Sigma _{\mu  a} \hat\partial_\mu}\,\hat
g~\ee^{-\Sigma _{\mu  a} \hat\partial_\mu}=
\hat\Omega_ a\,\hat g\,\hat\Omega_ a^\dagger \ ,
\label{ghattransf}\eeq
from which it follows that the gauge transformed operator in
(\ref{gaugeinvUp}) also satisfies the required twisted boundary condition
(\ref{constraintA_op}). Let us rewrite (\ref{ghattransf})
in terms of fields on the noncommutative torus $\IT_\Theta^D$. For this, we
expand the Weyl operator $\hat{g}$ using the basis $\hat{\Delta} (x)$ as
\beq
\hat{g} = \int \dd ^ D x ~ \hat{\Delta} (x) \otimes g (x) \ .
\label{hatgdef2}\eeq
Due to (\ref{ghattransf}), the $p\times p$ star($\star$)-unitary matrix field
$g(x)$ is multi-valued and defines an adjoint section of the gauge bundle,
\beq
g(x+\Sigma _{\mu a}\,\hat{\mu})=
\Omega_ a(x)\star g(x)\star\Omega_ a(x)^\dag \ .
\label{gstransf}\eeq

Let us now consider star($\star'$)-gauge invariant observables
in the U$(\tilde p_0)$ theory. Since the fields of this theory are
single-valued functions on the torus, we may use the construction of
Sections~\ref{observables} and \ref{nctorus}. Given the parallel
transport operators
\beq
{\cal U} ' (x';C) =
{\rm P}\,
\exp _{\star '} \left({i\int\limits_C
\dd \xi ^\mu ~{\cal A} ' _\mu (x'  + \xi) }\right) \ ,
\label{parallel_U1}
\eeq
where notation is as in Section~\ref{observables}, we may define a
star($\star'$)-gauge invariant observable by
\beq
{\cal O} (C) =
\int \dd ^D x ' ~\tr_{(\tilde p_0)}\Bigl({\cal U} ' (x '; C)\star ' S_v '
(x')\Bigr) \ ,
\eeq
where
\beq
S_v ' (x') = \ee ^{i k' _\mu x' _\mu}\,\id_{\tilde p_0} \ .
\eeq
The total momentum $k_\mu'$ is quantized as
\beq
k_\mu'=2\pi \left(\Sigma^{'-1}\right)_{ a\mu}\,m_ a '
\label{momprimeconstr}\eeq
while the constraint on the relative separation vector between the two ends of
the contour $C$ is
\beq
v_\mu =  \theta ' _{\mu \nu}\,k _\nu'  +\Sigma'_{\mu a}\,n'_ a \ ,
\label{vprimeconstr}\eeq
for some integer-valued vectors $m_ a'$ and $n_ a'$. We now rewrite the above
observables in terms of Weyl operators. Introducing
\beqa
\hat{U}'(C)& =& {\rm P}\,
\exp\left({\int\limits_C  \dd\xi^\mu~\left( \hat{\nabla} _\mu ^{(0)} +
i \, \hat{\cal A}_\mu\right) }\right) \ , \n
\hat{D}'(C) &= & {\rm P}\, \exp \left({ \int\limits_C  \dd\xi^\mu~
\hat{\nabla} _\mu ^{(0)} }\right) \ ,
\label{UDprime}\eeqa
we can define a star($\star'$)-gauge invariant observable as in (\ref{opendef})
by
\beq
{\cal O} (C) =
 \Tr '\,\tr_{(\tilde p_0)}\l \hat{U}'(C) \hat{D}'(C)^{\dag} \,\hat{S}_v \r \ ,
\label{defcalO_U1}
\eeq
where the unitary operator $\hat{S}_v$ may be expanded as
\beq
\hat{S}_v = \int \dd ^ D x ' ~\hat{\Delta} ' (x ')\otimes S_v ' (x ' ) \ ,
\label{hatSvprime}\eeq
and it satisfies
\beq
\hat{S}_v\,\hat{\Delta} ' (x')\,\hat{S}_v ^\dag
= \hat{\Delta}'(x ' + v)\otimes\id_{\tilde p_0} \ .
\eeq

Let us now see how to interpret the operator (\ref{defcalO_U1})
as a sensible quantity also in the Morita equivalent U($p$) theory.
We first define a parallel transport operator ${\cal U}(x;C)$ in the U($p$)
theory using Eqs.~(\ref{cU(C)}) and (\ref{cmproduct}). Note that $\hat{U}(C) =
\hat{U} '(C)$. The observable defined in (\ref{defcalO_U1}) can then be
rewritten as
\beq
{\cal O} (C) =\Tr\,\tr_{(p)}\biggl[\left( \hat{U}(C) \hat{D} (C)^{\dag}\right)
\left(\hat{D} (C) \hat{D}'(C)^{\dag}\right)\,\hat{S}_v \biggr] \ .
\label{defcalO_Up}
\eeq
We now expand the operator $\hat{S}_v$ in terms of the basis
$\hat{\Delta} (x)$ similarly to (\ref{hatgdef2}) and define
the $p\times p$ star($\star$)-unitary matrix field $S_v (x)$ as an adjoint
section. Then, using (\ref{cU(C)}), (\ref{UDprime}) and (\ref{defnabla}),
we can write the quantity (\ref{defcalO_Up}) as
\beq
{\cal O} (C) =\int \dd ^D x
{}~\tr_{(p)}\Bigl({\cal U}(x;C) \star S_v^{(0)}(x)\Bigr) \ ,
\label{OCSv0}\eeq
where we have defined the $p\times p$ star($\star$)-unitary matrix field
\beq
S_v^{(0)}(x)=\left[{\rm P}\,
\exp _{\star} \left( {i\int\limits_C \dd\xi^\mu~A_\mu ^{(0)}
(x+\xi)}\right)\right]^{*} \star S_v (x) \ .
\label{Sv0def}\eeq
By construction, (\ref{OCSv0}) is star($\star$)-gauge invariant, which can also
be checked by recalling the star($\star$)-gauge transformation
(\ref{gaugecovariant}) of the parallel transport operator ${\cal U}(x;C)$ and
the translation-generating property of the field (\ref{Sv0def}),
\beq
S_v^{(0)}(x)\star g(x)\star S_v^{(0)}(x)^\dag = g(x+v)  \ ,
\label{S0property}
\eeq
where $g(x)$ is an arbitrary adjoint section as in (\ref{gstransf}).
Note that the integrand of (\ref{OCSv0}) is a single-valued function because of
the boundary conditions
\beqa
{\cal U}(x+\Sigma _{\mu a}\,\hat{\mu};C)
&=& \Omega_ a(x)\star{\cal U}(x;C) \star\Omega_ a (x+v)^\dag \ , \\
S_v^{(0)}(x+\Sigma _{\mu a}\,\hat\mu)&=&\Omega_ a(x+v)\star
S_v^{(0)}(x)\star\Omega_ a(x)^\dag \ ,
\label{Sv0props}\eeqa
where the relation (\ref{Sv0props}) is required for consistency of the
equations (\ref{S0property}) and (\ref{gstransf}).

It follows that star($\star '$)-gauge invariant observables of the U($\tilde
p_0$) theory can be interpreted as star($\star$)-gauge invariant observables of
the U($p$) theory. As in Section~\ref{observables},
we can show that the field $S_v^{(0)}(x)$ satisfying (\ref{S0property}) on the
noncommutative torus $\IT_\Theta^D$ is in fact unique,
so that the Morita equivalence relation provides a one-to-one
mapping between the quantum correlation functions of the two noncommutative
Yang-Mills theories. An instructive example which shows how this correspondence
works is the $\theta = 0$ case. Then, one finds that Polyakov lines of
commutative gauge theory map to open loops in the Morita equivalent
noncommutative gauge theory.

The observables (\ref{OCSv0},\ref{Sv0def}) are the appropriate modifications of
the observables constructed in Sections~\ref{observables} and \ref{nctorus} to
the case of multi-valued gauge fields on the noncommutative torus. We remark
that the observables constructed here using the formalism of Weyl operators
differ from the noncommutative holonomy operators constructed recently in
\cite{Alekseev} which are star-gauge invariant without the need of introducing
the operator trace Tr, but which do not immediately generalize to gauge field
configurations of non-vanishing magnetic flux. The present noncommutative
Wilson lines are defined for multi-valued gauge fields, and moreover, as we
will see in Section~\ref{hopping}, they are the ones which arise in the
effective actions for charges propagating in the background of noncommutative
gauge fields.

\section{Lattice regularization}
\label{lattice}
\setcounter{equation}{0}

In this Section we will consider a lattice regularization of
noncommutative field theories. Through a general analysis of the noncommutative
algebra generated by the spacetime coordinates,
we will find that the discretization of the spacetime
inevitably requires that it be compact in some restricted
way depending on the noncommutativity.
Due to this restriction, the commutative limit is not commutable
with the continuum limit, which demonstrates the UV/IR mixing property of
noncommutative field theories at a fully nonperturbative level.
We will show that the discrete formulation allows a nonperturbative
regularization of generic noncommutative field theories with
single-valued fields. Given the Morita equivalence
property discussed in the previous Section,
this means that noncommutative Yang-Mills theories with gauge fields obeying
twisted boundary conditions can also be regularized
in terms of the Morita equivalent Yang-Mills theories with fields satisfying
periodic boundary conditions, insofar as defining regularized correlation
functions of star-gauge invariant observables are concerned.
In this way, we obtain a nonperturbative
definition of noncommutative Yang-Mills theories in
arbitrary even dimensions with multi-valued or single-valued gauge fields.

\subsection{General construction}
\label{general_construction}

To describe the lattice regularization of noncommutative Yang-Mills theory,
there are some subtle technical points concerning the transcription of the
continuum theory onto a lattice that we first need to address.
To this end, let us go back to the construction of Section~\ref{scalar}. To
define a lattice field theory, we restrict the spacetime points to
$x_\mu\in\epsilon\,\IZ$, where $\epsilon$ is the lattice spacing.
It follows that the lattice momentum must be identified
under the shift
\beq
k _\mu \mapsto k _\mu + \frac{2\pi}{\epsilon}\,\delta _{\mu\nu} \ ,
{}~~~~~~\nu = 1,\dots , D \ .
\label{momentumshift}
\eeq
Correspondingly, there is the operator identity
\beq
\ee^{ i (k_\mu  + \frac{2 \pi}{\epsilon}\,\delta _{\mu\nu}) \hat{x}_\mu}
= \ee^{ i  k_\mu  \hat{x}_\mu} \ ,~~~~~~\nu = 1,\dots , D \ .
\label{latticerestriction}
\eeq
By acting with the operator $\ee^{- i  k_\mu  \hat{x}_\mu}$ on both sides of
(\ref{latticerestriction}), we find that there is also the operator identity
\beq
\ee^{2 \pi i \hat{x}_\mu /\epsilon} = \id \ ,
{}~~~~~~\mu = 1,\dots , D \ ,
\label{latticeidentity}
\eeq
and, moreover, that the momentum $k_\mu$ is quantized according to
\beq
\theta _{\mu\nu} k_\nu\in2 \epsilon\,\IZ \ .
\label{latticemomentum}
\eeq
For $k_\mu$ satisfying (\ref{latticemomentum}), the commutation relation
\beq
\ee^{ik_\mu\hat x_\mu}~\ee^{2 \pi i \hat{x}_\nu /\epsilon}=\ee ^{2 \pi
i\theta_{\mu\nu} k_\mu / \epsilon }~
\ee^{2 \pi i \hat{x}_\nu /\epsilon}~\ee ^{i k_\mu \hat{x}_\mu} \ ,
{}~~~~~~\mu = 1,\dots , D
\label{momcommutator}
\eeq
is compatible with the identity (\ref{latticeidentity}).
On the other hand, the compatibility of the commutation relation
\beq
\ee ^{v _\nu   \hat{\del}_\nu}~
\ee^{2\pi i \hat{x}_\mu/\epsilon}~
\ee ^{-v  _\nu  \hat{\del} _\nu} =
\ee^{2\pi i (\hat{x}_\mu + v_\mu )/\epsilon}
\eeq
with the identity (\ref{latticeidentity}) requires $v_\mu\in\epsilon\,\IZ$.
Thus we are led to use instead the lattice shift operator
\beq
\hat{D}_\mu = \ee ^{\epsilon\,\hat{\del} _\mu}  \ ,
\label{latticeshiftop}
\eeq
but not the derivation $\hat{\del} _\mu$ itself anymore.
The restriction to the lattice shift operator
(\ref{latticeshiftop}) is standard in lattice field theory, whereas
the restriction (\ref{latticemomentum}) on the lattice momentum
simply disappears in the commutative case $\theta_{\mu\nu}=0$
and is quite characteristic of the noncommutative geometry.

The momentum quantization (\ref{latticemomentum})
requires that the spacetime be compactified in some restricted way.
Let us consider the case when the fields have the periodic boundary conditions
\beq
\phi ( x  + \Sigma _{\mu a}\,\hat \mu) = \phi (x)  \ ,
{}~~~~~ a = 1, \dots , D  \ ,
\label{sigmacompactification}
\eeq
where the periods $\Sigma _{\mu a}$ are integer multiples of the
lattice spacing $\epsilon$. Because of (\ref{sigmacompactification}),
the momentum is quantized as
\beq
k _\mu = 2 \pi (\Sigma ^{-1})_{ a\mu} m _ a \ ,
{}~~~~~~m _ a \in \IZ \ .
\label{momconstrlattice}
\eeq
The momentum periodicity (\ref{momentumshift}) due to
the lattice discretization $x _\mu\in\epsilon\,\IZ $ can be recast
in terms of the integers $m _ a$ as
\beq
m _ a \mapsto m _ a + \frac{1}{\epsilon}\,\Sigma _{\mu a} \ ,
{}~~~~~~\mu = 1,\dots , D \ .
\label{momentumperiodicity}
\eeq
Now the restriction on the momentum given by (\ref{latticemomentum}) implies
that there exists a $D \times D$ integer-valued matrix
$M _{\mu a}$ which satisfies
\beq
M _{\mu a} \Sigma _{\nu a} =
\frac{\pi}{\epsilon}\,\theta _{\mu\nu} \ .
\label{MST}
\eeq
We have therefore discovered the remarkable fact that
lattice regularization of noncommutative field theory
forces the spacetime to be compact. For fixed $M$, say $M = \id _D$,
the infrared cutoff disappears as $\frac{1}{\epsilon}$
in the continuum limit $\epsilon \rightarrow 0$.
Note that the commutative limit $\theta _{\mu\nu} \rightarrow 0$
does not commute with the continuum limit $\epsilon \rightarrow 0$.
Thus the UV/IR mixing discovered in perturbative
analyses of noncommutative field theories \cite{MRS} is demonstrated here
at a fully nonperturbative level. On the other hand, in order to obtain a
continuum spacetime of finite volume one has to send $M\to\infty$ as one takes
the $\epsilon \rightarrow 0$ limit. It is important to note
that for any given period matrix $\Sigma _{\mu a}$ and
noncommutativity parameter $\theta _{\mu\nu}$ in the continuum,
one can construct a family of lattice geometries
satisfying the restriction (\ref{MST}) and approaching the target continuum
theory in the $\epsilon \rightarrow 0$ limit.
This means that in spite of the restriction (\ref{MST})
for the regularized field theories, one can obtain the most general
noncommutative geometries parametrized by $\Sigma _{\mu a}$ and $\theta
_{\mu\nu}$
in the continuum limit.

Due to the momentum quantization (\ref{momconstrlattice}),
we are led to use the coordinate operators
\beq
\hat Z_ a = \ee ^{2 \pi i (\Sigma ^{-1}) _{ a\mu} \hat{x}_\mu } \ ,
\label{Zlattice}
\eeq
but not the $\hat{x} _\mu$ themselves anymore.
The commutation relations of the operators (\ref{Zlattice})
and (\ref{latticeshiftop}) are
\beqa
\label{Zlatticecomm}
\hat Z_ a\hat Z_ b
&=& \ee^{- 2 \pi i \Theta_{ a b}}\,\hat Z_ b\hat Z_ a \\
\hat D_\mu\,\hat Z_ a\,\hat D_\mu ^\dag &=&
\ee ^{2 \pi i \epsilon (\Sigma ^{-1}) _{ a\mu} }\, \hat Z_ a  \ ,
\label{Dlatticecomm}
\eeqa
where the dimensionless noncommutativity parameter
\beq
\Theta_{ a b} =
2\pi\left(\Sigma ^{-1}\right)_{ a\mu}\,\theta_{\mu\nu}
\left(\Sigma ^{-1}\right)_{ b\nu}
\label{Thetalattice}
\eeq
is necessarily rational-valued on the lattice,
since the restriction (\ref{MST}) implies that
\beq
M_{\mu a} = \frac1{2\epsilon}\,\Sigma_{\mu b} \Theta_{ b a}
\label{MSTdimless}
\eeq
is an integer-valued matrix. Because of (\ref{latticeidentity}), we also have
the identity
\beq
\prod _{ a =1} ^{D}
\left(\hat Z_ a \right)^{\frac{1}{\epsilon}\,\Sigma_{\mu a}}
= \exp \left[  \pi i \sum _{ a <  b} \Sigma _{\mu a}
\Theta _{ a  b} \Sigma _{\mu  b}  \right] \cdot
\id _D \ ,
{}~~~~~~\mu = 1,\dots , D \ .
\label{latticeidentity_Z}
\eeq
By construction, the identity (\ref{latticeidentity_Z})
is compatible with the commutation relations
(\ref{Zlatticecomm}) and (\ref{Dlatticecomm}).

We can define a map $\hat{\Delta}(x)$ between lattice fields and Weyl operators
by
\beq
\hat{\Delta} (x) =
\frac1{\left|\det \frac1\epsilon\, \Sigma\right|}\,\sum_{\vec{m}}
\left(\prod_{ a=1}^D\left(\hat Z _ a\right)^{m_ a}\right)
{}~\ee ^{- \pi i\sum_{ a< b}\Theta_{ a b}m_ a m_ b}~
\ee ^{- 2 \pi i (\Sigma ^{-1})_{ a\mu} m_ a x_\mu } \ ,
\label{map_lattice}
\eeq
where the sum goes over $\IZ ^D$ modulo the periodicity
(\ref{momentumperiodicity}) and $x_\mu$ is a point on the spacetime lattice
$\epsilon\,\IZ$ with periodicity (\ref{sigmacompactification}). Note that
\beq
\frac1{\left|\det \frac1\epsilon\, \Sigma\right|}\,\sum_{\vec{m}}
\ee ^{2 \pi i (\Sigma ^{-1})_{ a\mu} m_ a x_\mu }
= \delta _{x , 0 (\mbox{\scriptsize mod}\,\Sigma)} \ .
\eeq
As we did in (\ref{starproduct}), we can define the lattice star-product
using the map $\hat{\Delta} (x)$ defined in (\ref{map_lattice}).
Explicitly it can be given as
\beq
\phi_1(x)\star\phi_2(x)=
\sum_{y,z} K(x-y,x-z)\,\phi_1(y)\,\phi_2(z)~ \ ,
\label{latticestarprod}\eeq
where the sums go over the spacetime lattice points $\epsilon\,\IZ$
modulo the periodicity (\ref{sigmacompactification}).
The kernel $K$ in (\ref{latticestarprod}) is given by
\beq
K(y,z)
= \frac1{\left|\det \frac1\epsilon\, \Sigma\right|^2}\,\sum _{\vec{m},\vec{n}}
\exp\Bigl[ 2 \pi i (\Sigma  ^{-1})_{ a\mu}
( m_ ay_ \mu  + n_ az_ \mu)
+ i \pi \Theta _{ a b} m_ a n_ b\Bigr]  \ .
\label{latticekernel}
\eeq
The map (\ref{map_lattice}) and the star-product
(\ref{latticestarprod},\ref{latticekernel}) possess all the algebraic
properties that
their continuum counterparts have. If $2 \epsilon (\Sigma \Theta ) ^{-1} =
M^{-1}$ is an integer-valued matrix,
where $M$ is the integral matrix introduced in (\ref{MST}),
then the sums in (\ref{latticekernel}) can be done explicitly yielding
\beq
K(y,z)=  \frac1{\left|\det \frac1\epsilon\, \Sigma\right|}~\ee^{-
2 i (\theta^{-1})_{\mu\nu}y_\mu z_\nu} \ .
\label{latticekernelsimple}
\eeq
The corresponding formula (\ref{latticestarprod}) for the lattice star-product
is then analogous to the second expression in (\ref{starproduct}) for the
continuum star-product.

Let us now turn to the construction of field theories on a
discrete noncommutative torus.
In the case of a scalar lattice field $\phi(x)$, one can define the operator
\beq
\hat{\phi} = \sum_x\phi (x)\,\hat{\Delta} (x)
\eeq
where the sum runs over lattice points. We may then write down an action
\beq
S\left[\hat{\phi}\right] =
\Tr
\left( \frac{1}{2}\,\sum_\mu\left(\hat{D}_\mu\,\hat{\phi}\,\hat{D}_\mu^\dag -
 \hat{\phi}\right)^2 + \frac{1}{2}\,\hat{\phi} ^2
+ \frac{1}{4!}\,\hat{\phi} ^4 \right)  \ ,
\eeq
which leads to the usual lattice action of the field $\phi(x)$
with star-interaction term analogous to (\ref{four-int}).

A huge advantage of the lattice regularization is that it allows one to
construct finite dimensional representations of the noncommutative geometry.
This is apparent already at the level of the lattice operators
(\ref{map_lattice}). It is straightforward to compute that they obey the
commutation relations
\beq
\left[\hat\Delta(x)\,,\,\hat\Delta(y)\right]=\sum_z{\cal K}(x-z,y-z)
{}~\hat\Delta(z) \ ,
\label{DeltaLiealg}\eeq
where
\beq
{\cal K}(x,y)=\frac{2i}{\left|\det\frac1\epsilon\,\Sigma\right|^2}\,
\sum_{\vec m,\vec n}\sin\Bigl(\pi\Theta_{ab}m_an_b
\Bigr)~\ee^{2\pi i(\Sigma^{-1})_{a\mu}(m_ax_\mu+n_ay_\mu)} \ .
\label{calkernel}\eeq
The lattice operators $\hat\Delta(x)$ therefore generate a finite dimensional
Lie algebra of dimension $|\det\frac1\epsilon\,\Sigma|=2^D|\det M/\det\Theta|$.
In the lattice field theory, the operators $\hat\Delta(x)$ can in this way be
regarded as maps from the $|\det\frac1\epsilon\,\Sigma|$ lattice points to a
finite dimensional matrix representation, thereby illustrating how the degrees
of freedom are mapped into each other in the one-to-one correspondence between
Weyl operators and fields. Indeed, the algebra (\ref{Zlatticecomm}) can always
be represented by finite dimensional matrices which admit a tensor product
decomposition into blocks depending on the detailed forms of the rational
numbers $\Theta_{ab}$ (See \cite{AMNS} for some specific examples). We shall
return to this point in the next Section. The lattice regularization can
therefore also be thought of as an approximation to the algebra of Weyl
operators in the continuum by finite dimensional matrix algebras. This
constitutes the standard description of noncommutative lattices by
approximately finite algebras in noncommutative geometry \cite{NClattice}. The
rigorous definition of the continuum limit described above within such an
algebraic description of the noncommutative geometry is given in
Ref.~\cite{lls}.

\subsection{Noncommutative Yang-Mills theory on the lattice}
\label{latticeNCYM}

In order to construct a lattice regularization of noncommutative Yang-Mills
theory, we need to maintain star-gauge invariance on the lattice.
As in the case of ordinary lattice gauge theory \cite{Wilson},
this is achieved
by putting the gauge fields on the links of the lattice \cite{AMNS},
\beq
\hat{U}_\mu = \sum _x\hat{\Delta} (x) \otimes U_\mu (x) \ ,
\label{defUhat}
\eeq
where $\hat{U}_\mu$ is a unitary operator and $U_\mu(x)$
is a $p \times p$ matrix field on the lattice which is star-unitary,
\beq
U_\mu(x) \star U_\mu(x)^{\dag}=\id_p \ .
\label{starunitary_link}
\eeq
One can write an action
\beq
S =-\frac{1}{g^2}\,\sum_{\mu\neq \nu}\Tr\,\tr_{(p)}\left[ \hat{U}_\mu
\left(\hat{D}_\mu\,\hat{U}_\nu\,\hat{D}_\mu^\dag\right)
\left(\hat{D}_\nu\,\hat{U}_\mu ^\dag\,\hat{D}_\nu^\dag\right)
\hat{U}_\nu ^\dag \right]
\label{action}
\eeq
which is invariant under the transformation
\beq
\hat{U}_\mu \mapsto \hat{g}\,\hat{U} _\mu\,
\left(\hat{D}_\mu\,\hat{g}^\dag\,\hat{D}_\mu ^\dag\right) \ .
\eeq
The action (\ref{action}) can be written in terms of the lattice fields $U_\mu
(x)$ as
\beq
S = -\frac1{g^2}\,
\sum _x\sum_{\mu\neq\nu}\tr_{(p)}\Bigl[U_\mu (x) \star
U_\nu (x + \epsilon \hat{\mu}) \star
U_\mu (x + \epsilon \hat{\nu}) ^\dag \star
U_\nu (x) ^\dag\Bigr] \ ,
\label{latticeaction}
\eeq
which is invariant under the lattice star-gauge transformation
\beq
U_\mu (x)\mapsto g(x) \star U_\mu (x)
\star g(x+\epsilon \hat{\mu})^\dag  \ ,
\label{latticestargaugetr}
\eeq
where the gauge function $g(x)$ is defined by
\beq
\hat{g} = \sum _x\hat{\Delta} (x) \otimes g (x)
\label{defghat}
\eeq
and it is star-unitary, $g(x)\star g(x)^{\dag}=\id_p$.

This discrete version of noncommutative Yang-Mills theory has several benefits.
First of all, it provides a concrete definition of the quantum gauge theory
path integral in the continuum. The unitary operators (\ref{defUhat}) live in a
finite-dimensional operator algebra and are therefore elements of a compact
unitary group. Note that the trace $\Tr\,\tr_{(p)}$ which appears in
(\ref{action}) corresponds to the trace in the fundamental representation of
this unitary group. The lattice gauge theory path integral may then be defined
by the measure $\dd\hat U$ which is the Haar measure of
this compact Lie group that is invariant under the left and right actions
\beq
\hat U_\mu\mapsto\hat g\,\hat U_\mu~~~~~~,~~~~~~\hat U_\mu\mapsto\hat
U_\mu\,\hat g \ .
\label{Haardef}\eeq
In terms of the star-unitary lattice fields $U_\mu (x)$, the path integral
measure ${\cal D}U_\mu(x)$ is given by the Haar measure of the compact Lie
group which is formed by the $U_\mu(x)$ with multiplication given by the
star-product, and which is invariant under the left and right actions $U_\mu(x)
\mapsto g (x)\star U_\mu(x)$ and $U_\mu(x) \mapsto U_\mu(x)\star g (x)$. This
measure preserves star-gauge invariance. In the commutative limit $\theta
_{\mu\nu} = 0$, the discrete noncommutative Yang-Mills theory reduces to
ordinary lattice gauge theory \cite{Wilson}, including the path integral
measure because of the uniqueness of the Haar measure. The lattice
regularization also gives an approximation to the star-gauge symmetry group in
the continuum. An algebraic description of this continuum gauge group is given
in Ref.~\cite{lschaos}.

Let us now describe the discrete analogs of the star-gauge invariant
observables
introduced in Section~\ref{observables}. We first define the discrete analog of
the parallel transport operator ${\cal U}(x;C)$. We introduce an oriented
contour $C$ on the lattice specified by a collection of links,
\beq
C = \{ \mu_1 , \mu_2 , \dots , \mu_n  \} \ ,
\label{contour_lattice}
\eeq
where $\mu_j = \pm 1, \pm 2 , \dots,\pm D$.
We define $U _{-\mu} (x)= U_\mu (x - \epsilon \hat{\mu})^\dag$.
The parallel transport operator can then be defined as
\beq
{\cal U}(x;C)
= U_{\mu_1} (x) \star U_{\mu _2} (x+ \epsilon \hat{\mu}_1) \star \cdots
\star U_{\mu _n}\left(x+ \epsilon \sum _ {j=1}^{n-1}\hat{\mu}_j\right) \ .
\eeq
As in the continuum, in order to construct star-gauge invariant observables
out of the operator ${\cal U}(x;C)$, we need a star-unitary function $S_v (x)$
with the property (\ref{Svprop}) for arbitrary functions $g(x)$
on the periodic lattice, where
\beq
v=\epsilon\sum_{j=1}^n\hat\mu_j \ .
\eeq
One finds that a necessary and sufficient condition is given
again by (\ref{Svequ}) with the solution $S_v (x ) = \ee ^{i k _\mu  x_\mu
}\,\id_p$, where the loop momentum $k_\mu$ satisfies (\ref{vk_relation}).
In the present case, both $k$ and $v$ are quantized and periodic.
Modding out by the periodicity, there are only the same, finite number
of values that $k$ and $v$ can take. Therefore, it makes sense to ask whether
or not (\ref{vk_relation}) gives a one-to-one correspondence between
$k$ and $v$. The answer is affirmative if and only if
there exist $D \times D$ integer-valued matrices $J$ and $K$ which satisfy
\beq
\frac{1}{\epsilon}\,\Sigma\,J - 2 M K =  \id _D \ ,
\label{JKrel}
\eeq
where $M$ is the integral matrix introduced in (\ref{MST}).
If this condition is not met, then there exist loop separation vectors $v$ for
which there is no momentum $k$ satisfying (\ref{vk_relation}), and
for the other $v$ there is more than one value of $k$ satisfying
(\ref{vk_relation}). For example, let us consider the case with period matrix
$\frac{1}{\epsilon}\,\Sigma = L \id _D$.
If $L$ is an even integer, then (\ref{JKrel}) cannot be satisfied.
If $L$ is odd and $M ^{-1}$ is an integral matrix, then
one can satisfy (\ref{JKrel}) by taking
$J=\id_D$ and $K=\frac{L-1}{2}\,M^{-1}$.

Thus, for periodic gauge fields,
one can construct a lattice regularization of noncommutative
Yang-Mills theory with arbitrary gauge group rank $p$, period $\Sigma$,
and noncommutativity parameter $\theta$. Due to Morita equivalence in the
continuum, this means that we have a nonperturbative formulation of the general
class of (twisted) continuum gauge theories that were described in
Section~\ref{GTderivation1}. In this regard, it is important that Morita
equivalence also
holds at the level of observables as we discussed in
Section~\ref{observables_morita}. The problem with defining a discrete
noncommutative U$(p)$ gauge theory
with multi-valued gauge fields directly is that one immediately encounters an
obstacle to constructing a background abelian gauge field
(\ref{con_u_0_higher}) on the lattice. As we discussed above, the relation
(\ref{latticemomentum}) imposes a quantization constraint on the allowed
momenta in the lattice discretization of noncommutative geometry.
This would imply that the background tensor field $F$ be
quantized proportionally to $\theta^{-1}$,
and so the configuration (\ref{con_u_0_higher})
cannot be constructed on the lattice in general.
However, as we have discussed, we circumvent this difficulty
by using the noncommutative lattice gauge theory with single-valued gauge
fields and Morita equivalence in the continuum to obtain a non-perturbative
definition of all noncommutative Yang-Mills theories associated with constant
curvature vector bundles over tori. Other problems with discretizing
noncommutative geometry, such as the construction of an appropriate lattice
Dirac operator, are discussed in Ref.~\cite{gslattice}.

\section{Explicit realizations of discrete noncommutative gauge theory}
\label{explicit}
\setcounter{equation}{0}

The construction of the previous Section has been quite general, and we will
now describe some concrete examples of discrete noncommutative Yang-Mills
theory. Within the lattice formalism, we will demonstrate the Morita
equivalence between commutative U($p$) lattice gauge theory with fields obeying
twisted boundary
conditions and a noncommutative U($\tilde p_0$) lattice gauge theory with
fields obeying periodic boundary conditions. Using this property, we will
further show that noncommutative gauge theory can be regularized by means of
{\em commutative} lattice gauge theory with 't~Hooft flux. As a special case,
this construction includes a previous proposal for a concrete definition of
noncommutative Yang-Mills theory using large $N$ reduced models \cite{AIIKKT}.
This will further lead to the explicit finite dimensional matrix
representations of the noncommutative torus that were discussed in the previous
Section.

\subsection{Discrete Morita equivalence}
\label{simpleMorita}

In Section~\ref{latticeNCYM}, we have shown that noncommutative Yang-Mills
theories can be nonperturbatively regularized.
The regularized theory is described by star-unitary gauge fields
(\ref{starunitary_link})
on a lattice and the action (\ref{latticeaction})
is defined in terms of lattice star-products.
Although the theory is explicitly given in terms of a finite number
of degrees of freedom, as it stands
it is not very suitable for practical purposes,
say for numerical studies such as Monte Carlo simulations.
In this Subsection, we show further that noncommutative Yang-Mills theory
can be nonperturbatively regularized by
means of {\em commutative} lattice gauge theories
with multi-valued gauge fields \cite{tHooft}.

For this, we consider the lattice analog of
Morita equivalence which we derived in Section~\ref{GTderivation2}.
As we discussed in Section~\ref{latticeNCYM},
there is a technical obstruction
to constructing noncommutative Yang-Mills theory
directly on the lattice for gauge fields of non-vanishing topological charge.
We recall that the only obstacle was that the constant
abelian background gauge field (\ref{con_u_0_higher}) does not
generally have a momentum compatible
with the restriction (\ref{latticemomentum}).
This obstacle disappears, of course, for the commutative case
$\theta _{\mu\nu} = 0$. What we will prove in the lattice formulation is
the Morita equivalence between {\em commutative} Yang-Mills theory
with gauge fields obeying twisted boundary conditions and noncommutative
Yang-Mills theory with periodic gauge fields, in arbitrary even dimension
$D=2d$.
In fact, we will find that, for a given noncommutative Yang-Mills theory
with arbitrary period matrix $\Sigma$ and deformation parameter $\Theta$ in the
continuum, we can construct a family of
commutative U($p$) lattice gauge theories with 't~Hooft flux whose sequence of
Morita equivalent noncommutative theories converges to the target
noncommutative field theory in the continuum limit.
When $\Theta$ is irrational, the rank $p$ of the gauge group of
the commutative theory must be sent to infinity
as one takes the continuum limit.
Combining this with the continuum Morita equivalence which connects
two noncommutative Yang-Mills theories with
periodic and twisted boundary conditions on the gauge fields,
we will find that continuum noncommutative Yang-Mills theories,
with either rational or irrational noncommutativity parameters and
with or without twists in the boundary conditions on the gauge fields,
can be regularized by means of {\em commutative} lattice gauge theory with
't~Hooft flux.

We start with a commutative U$(p)$ lattice gauge theory with fields
obeying twisted boundary conditions. The action is
\beq
S =-\frac{1}{g ^2}\,\sum _{x}  \sum_{\mu \neq \nu}\tr_{(p)}\Bigl[U_\mu (x)\,
U_\nu (x + \epsilon \hat{\mu})\,U_\mu (x + \epsilon \hat{\nu}) ^\dag\,
U_\nu (x) ^\dag\Bigr] \ ,
\label{commutativeaction}
\eeq
where $U_\mu (x)$ are U($p$)
gauge fields satisfying the twisted boundary conditions
\beq
U_\mu (x+ \Sigma _{\nu a}\,\hat{\nu}) =
\Omega _ a(x)\,U_\mu (x)\,\Omega _ a
(x+ \epsilon \hat{\mu})^\dag
\label{bc}
\eeq
with period matrix $\Sigma$. The transition functions $\Omega_ a (x)$
are U($p$) matrices which, for consistency of the constraints (\ref{bc}), must
satisfy the cocycle condition
\beq
\Omega _ a (x + \Sigma _{\mu b}\,\hat{\mu}) ~ \Omega _ b (x)
= {\cal Z}_{ a b}~ \Omega _ b (x+ \Sigma _{\mu a}\,\hat{\mu}) ~
\Omega _ a (x) \,,
\label{cocycleD_lattice}
\eeq
where ${\cal Z}_{ a b}=\ee^{2\pi i\gamma_{ a b}/p}\in\IZ_p$.
The antisymmetric matrix $\gamma$ has elements $\gamma_{ab}\in\IZ$
representing the 't~Hooft fluxes. As an explicit form of the transition
functions, let us take the gauge choice
\beq
\Omega_ a(x)=
1  \otimes \Gamma_ a \ ,
\label{con_omega_explicit}
\eeq
where $\Gamma_ a$ are the SU$(p)$ twist eaters obeying the Weyl-'t~Hooft
commutation relations (\ref{GammaalgD_higher}). In order to satisfy
(\ref{cocycleD_lattice}) we must have $Q = \gamma$. Note that the gauge choice
(\ref{con_omega_explicit}) corresponds to setting $\alpha=0$ in the continuum
expression (\ref{con_omega_explicit_higher}). The reason for this is that, as
we discussed in Section~\ref{GTderivation1}, keeping both $\alpha$ and $\gamma$
non-zero in the commutative case is redundant. In other words, the abelian
magnetic flux of a U$(p)$ gauge field, which plays a very important role in
Morita equivalences of
noncommutative gauge theories~\cite{Morita}, arises in the
commutative case only via the corresponding 't~Hooft flux~\cite{LPR}.
A similar consideration with non-vanishing $\alpha$
would lead us to the same final results. Due to the constraint (\ref{bc}), the
gauge theory can be expressed in terms of lattice gauge fields $U_\mu (x)$ with
$x_\mu$ lying in a unit cell of period $\Sigma$.

We will now show that the lattice field theory (\ref{commutativeaction})
with the constraint (\ref{bc}) is equivalent to a noncommutative U($\tilde
p_0$) lattice gauge theory with fields obeying periodic boundary conditions and
the reduced rank $\tilde p_0$ defined by (\ref{tildep0def}).
For this, we rewrite the lattice field theory (\ref{commutativeaction})
in terms of (finite dimensional) commutative operators. We write
\beqa
\label{defUhat_Omegahat}
\hat{U}_\mu &=& \sum _x\hat{\Delta} (x) \otimes U_\mu (x) \\
\hat{\Omega} _ a &=& \sum_x\hat{\Delta} (x) \otimes \Omega _ a (x) \ ,
\label{hatOmegalattice}\eeqa
where the map $\hat\Delta(x)$ is defined by (\ref{map_lattice}) and the
operator $\hat\Omega_ a$ is given explicitly by
\beq
\hat{\Omega} _ a =
\id \otimes\Gamma_ a \ .
\eeq
The action (\ref{commutativeaction}) can then be written as
\beq
S =-\frac{1}{g^2}\,\sum_{\mu\neq \nu}\Tr\,\tr_{(p)}\left[ \hat{U}_\mu
\left(\hat{D}_\mu\,\hat{U}_\nu\,\hat{D}_\mu^\dag\right)
\left(\hat{D}_\nu\,\hat{U}_\mu ^\dag\,\hat{D}_\nu^\dag\right)
\hat{U}_\nu ^\dag \right] \ ,
\label{opactionlattice}\eeq
where $\hat D_\mu$ are the lattice shift operators (\ref{latticeshiftop}), and
the constraint (\ref{bc}) becomes
\beq
\left(\hat{D} _\nu\right)^{\frac1{\epsilon}\,\Sigma_{\nu a}}
\,\hat{U} _\mu\,\left(\hat{D} _\nu^\dag\right)
^{\frac1{\epsilon}\,\Sigma_{\nu a}}
 = \hat{\Omega} _ a\,\hat{U} _\mu\,\hat{\Omega} _ a ^ \dag \ .
\label{opconstrlattice}\eeq
Proceeding exactly as in Section 3.1, one finds that
the general solution to the constraint (\ref{opconstrlattice}) is given by
\beq
\hat{U}_\mu =\sum_{\vec{m}}
\,\left(\prod_{ a=1}^D\left(\hat Z_ a '\right)^{m_ a}\right)
{}~\ee^{ - \pi i\sum_{ a< b}\Theta ' _{ a b} \,m_ a m_ b}
\otimes u_\mu(\vec{m}) \ ,
\label{UiZD}
\eeq
where $u_\mu(\vec{m})$ is a $\tilde p_0 \times \tilde p_0$ matrix and
\beq
\hat Z'_ a =\ee^{2\pi i\bigl(\Sigma ^{'-1}\bigr)_{ a\mu}\,\hat{x}_\mu}
\otimes\prod_{ b=1}^D(\Gamma_ b)^{B_{ a b}} \ .
\label{Ziansatz_U}
\eeq
The $D\times D$ matrices $\Sigma '$ and $\Theta '$ are given by
\beqa
\label{Sigmaprime}
\Sigma ' &=&  \Sigma\,\tilde{P} \\
\Theta ' &=&  - \tilde{P} ^{-1} B^\top \ ,
\label{Thetaprime}
\eeqa
where the integral matrices $B$ and $\tilde P$ are defined in
Section~\ref{GTderivation2}. Because of their dependence on the twist eating
solutions, the operators (\ref{Ziansatz_U}) obey the commutation relations
\beqa
\hat{Z}_ a '\hat{Z}_ b '
&=&  \ee ^{- 2 \pi i \Theta ' _{ a b}}\,\hat{Z}_ b '\hat{Z}_ a '  \n
\hat D_\mu\,\hat Z_ a '\,\hat D_\mu ^\dag &=&\ee^{ 2\pi i \,\bigl(
\Sigma ^{' -1}\bigr)_{ a\mu}}\,\hat Z_ a ' \ .
\label{Zprimelatticecomm}\eeqa
The commuation relations (\ref{Zprimelatticecomm}) are the same as
(\ref{Zlatticecomm}) and (\ref{Dlatticecomm})
with the replacements $\hat Z \to \hat Z '$, $\Theta \to \Theta ' $ and
$\Sigma \to \Sigma ' $. The two matrices $\Sigma '$ and $\Theta '$
must satisfy the general constraint (\ref{MSTdimless}), which implies that
$M=- \frac{1}{2 \, \epsilon}\,\Sigma   \, B ^\top$
must be an integer-valued matrix. The sum over $\vec{m}$ in (\ref{UiZD})
can then be taken over $\IZ ^D$ modulo the periodicity
$m_ a \sim m_ a + \frac1\epsilon\,\Sigma ' _{\mu a}$ with $\mu = 1,\dots , D$.

We now introduce a corresponding map $\hat{\Delta} ' (x ')$ analogously to
(\ref{map_lattice}) and decompose the operator $\hat{U}_\mu$ in this new basis
as
\beq
\hat{U}_\mu = \sum _{x '}\hat{\Delta} ' (x ')\otimes U_\mu ' (x') \ .
\label{newexpansion}
\eeq
Substituting (\ref{newexpansion}) into (\ref{opactionlattice}), we arrive at
the action
\beq
S = -\frac{1}{g'^2}\,\sum _{ x '}\sum_{\mu\neq\nu}\tr_{(\tilde
p_0)}\Bigl[U_\mu  ' (x ') \star'U_\nu  ' (x ' + \epsilon \hat{\mu}) \star'
U_\mu ' (x ' + \epsilon \hat{\nu}) ^\dag\star'U_\nu  ' (x ') ^\dag\Bigr]
\label{latticeaction2}
\eeq
where
\beq
g'^2 =\frac{p}{\tilde p_0} \,g^2 \ .
\label{gprimelattice}\eeq
This shows that an ordinary U($p$) lattice gauge theory with multi-valued gauge
fields is equivalent to noncommutative U($\tilde p_0$) lattice gauge theory
with single-valued gauge fields and deformation parameter matrix
(\ref{Thetaprime}).

We now show that for any noncommutative Yang-Mills theory with given $\Theta '$
and $\Sigma '$ in the continuum and with gauge group U($\tilde{p}_0$), we can
construct a family of commutative U($p$) lattice gauge theories with
multi-valued gauge fields whose sequence of Morita equivalent theories
converges to the target noncommutative gauge theory in the continuum limit. Let
us first consider the case where the noncommutativity parameter matrix $\Theta
'$ given in the continuum is rational-valued. By using the SL($D,\IZ$)
transformation
$\Theta ' \mapsto \Lambda\,\Theta '\,\Lambda ^\top$, where
$\Lambda \in \mbox{SL}(D,\IZ)$,
we can rotate $\Theta '$ into the canonical skew-diagonal form
\beq
\Theta '  =\pmatrix{0&- \vartheta_1 '
& & & \cr  \vartheta_1 ' &0& & & \cr &
&\ddots& & \cr & & &0&-\vartheta_d '  \cr & &
& \vartheta_d '  &0\cr}  \ ,
\label{Thetadiag}
\eeq
where the skew-eigenvalues $\vartheta _i '$ are also rational numbers.
Let us denote them by $\vartheta _i ' = - \frac{b_i}{\tilde{p}_i} $,
where $b_i$ and $\tilde{p}_i$ are relatively prime integers for each
$i=1,\dots,d$. Then there exist integers $a_i$ and $\tilde{q}_i$ which satisfy
(\ref{apbqpi}). We can define the lattice rank $p$ by (\ref{tildep0def})
and the twist matrix $Q$ by (\ref{QtildeQ}). Since the matrix $\tilde{P}$ is
invertible, we can approximate the period $\Sigma '$ to arbitrary precision by
the matrix $\Sigma\,\tilde{P}$ in the continuum limit $\epsilon\to0$,
while keeping fixed the lattice period $\frac1{\epsilon}\,\Sigma$ to an
integer-valued matrix.

When $\Theta '$ is irrational, one has to consider a rational-valued
deformation parameter matrix $\Theta_\epsilon'$ for each lattice spacing
$\epsilon$,
which converges to the given irrational numbers $\Theta_\epsilon'\to\Theta'$ as
$\epsilon\to0$. This means that the corresponding integers $b_{\epsilon,i}$ and
$\tilde{p}_{\epsilon,i}$ should go to infinity in the continuum limit.
Thus,  the corresponding $p_\epsilon$'s given by
(\ref{tildep0def}) should also go to infinity.
We conclude that, in order to define a continuum noncommutative gauge theory
with irrational noncommutativity parameters,
the rank $p$ of the gauge group of the approximating commutative lattice gauge
theory should be sent to infinity as one takes the continuum limit.
Note that if one takes the 't Hooft (planar) limit by
fixing $\lambda = g^2\,p$ as $p\to\infty$, then
the coupling constant $g '$ of the Morita equivalent
noncommutative U($\tilde{p}_0$) gauge theory
is fixed according to (\ref{gprimelattice}).
Therefore, in the 't~Hooft limit, the lattice spacing $\epsilon$ should be
taken to zero only {\it after} one takes the limit $p\to\infty$ and the limit
should be accompanied by an appropriate renormalization of
the 't~Hooft coupling constant $\lambda$.
However, in this limit the dimensionful noncommutativity parameters $\theta '$
inevitably diverge. On the other hand, in order to have finite $\theta '$ one
should take a ``double scaling limit'' by sending $\epsilon\to0$
together with the $p\rightarrow \infty$ limit in a correlated way.
In this limit, non-planar Feynman diagrams survive in addition to the planar
diagrams. Thus, although we have to take the large $p$ limit in order to
obtain irrational $\Theta '$, this limit should not be confused with the
't~Hooft limit in general.

\subsection{Twisted Eguchi-Kawai model}
\label{TEK_model}

We will now consider the construction of the previous Subsection in the special
case where the period matrix is $\Sigma=\epsilon\id_D$.
A one-site U($p$) lattice gauge theory is just the Eguchi-Kawai model
\cite{EK}, and the fact that the boundary conditions are twisted as in
(\ref{bc})
means that it is actually the twisted Eguchi-Kawai model \cite{GO}.
To see this explicitly, we reduce the action (\ref{commutativeaction})
to a single point $x=0$ by using the constraints (\ref{bc}) to get
\beq
U_\mu(\epsilon\,\delta_{\nu a}\,\hat\nu)=\Gamma_a\,U_\mu(0)\,(\Gamma_a)^\dagger
\ .
\label{Ured}\eeq
Substituting (\ref{Ured}) into (\ref{commutativeaction}) at $x=0$ and using the
commutation relations (\ref{GammaalgD_higher}), we arrive at the action
\beq
S=-\frac{1}{g^2}\,
\,\sum_{\mu\neq\nu} {\cal  Z}_{\mu\nu}~\tr_{(p)}\Bigl(V_\mu\,V_\nu\,
V_\mu^\dag\,V_\nu^\dag\Bigr)
\label{EKaction}
\eeq
where $V_\mu=U_\mu (0) \,\Gamma_\mu$, $\mu=1,\dots,D$, are
$p\times p$ unitary matrices. This is the action of the twisted Eguchi-Kawai
model, where the phase factor ${\cal Z}_{\mu\nu}=\ee^{2\pi i\gamma_{\mu\nu}/p}$
is called the ``twist''. Thus, the recent proposal \cite{AIIKKT} that the
twisted large $N$ reduced model
serves as a concrete definition of noncommutative Yang-Mills theory
can be interpreted as the simplest example of Morita equivalence.
The possibility of such an interpretation has also been suggested
in Ref. \cite{HI}.

Conversely, one can derive the explicit map from matrices in the twisted
Eguchi-Kawai model to U($\tilde p_0$) gauge fields on the noncommutative torus
using the present formalism. For this, we decompose the matrices $V_\mu$ of the
action (\ref{EKaction}) using the SU($p$) twist eaters $\Gamma_\mu $ as
\beq
V_\mu  =U_\mu\,\Gamma_\mu \  .
\label{UvsU}
\eeq
We then use the twisted boundary conditions (\ref{bc}) to generate from the
$p\times p$ unitary matrix $U_\mu(0)=U_\mu$ a commutative U$(p)$ lattice gauge
field $U_\mu(x)$, and introduce the operator $\hat U_\mu$ by
(\ref{defUhat_Omegahat}). Then, using the expansion (\ref{newexpansion}) and
remembering that the original field theory is reduced to a point $x=0$, the
U$(\tilde p_0)$ gauge field $U_\mu'(x')$ on the noncommutative torus can be
given as
\beqa
U_\mu'(x')&=&\Tr'\Bigl(\hat\Delta'(x')\,\hat U_\mu\Bigr)\n
&=&\frac p{\tilde p_0}\,\sum_{\vec{m}}\ee^{ \pi i \sum_{\mu<\nu}
\Theta ' _{\mu\nu} \,m_\mu m_\nu}{}~\ee ^{- 2 \pi i
\left(\Sigma ^{' -1}\right) _{\mu\nu}\,m_\mu\,x_\nu'} \n
& &\times~\tr_{(\tilde p_1\otimes\cdots\otimes\tilde
p_d)}\left[V_\mu\,(\Gamma_\mu)^\dag\,\prod_{\nu=1}^D\left(\prod_{\rho=1}^D
(\Gamma_\rho)^{B_{\nu\rho}}\right)^{m_\nu}\right] \ .
\label{EKmap}\eeqa
Eq.~(\ref{EKmap}) gives the mapping between the $p\times p$ unitary matrices
$V_\mu$ of the twisted Eguchi-Kawai model and noncommutative U$(\tilde p_0)$
lattice gauge fields $U_\mu'(x')$ with noncommutativity parameter matrix
(\ref{Thetaprime}) and period matrix (\ref{Sigmaprime}).
This shows explicitly how, via Morita equivalence,
large $N$ reduced gauge theories serve as a nonperturbative formulation of
noncommutative Yang-Mills theory.

As in the case of general periods $\Sigma$ which we discussed in the previous
Subsection, one can reproduce arbitrary dimensionless noncommutativity
parameters $\Theta '$ in the continuum limit. However, since $\Sigma =\epsilon
\id _D$ in the present case, the induced period matrix $\Sigma '$ in
(\ref{Sigmaprime}) is not the most general one. For example, in $D=2$
dimensions, one can obtain only a square torus, i.e. $\Sigma ' = \ell\,\id _2$,
as the continuum spacetime. The rank $p$ of the gauge group of the reduced
model
must be sent to infinity even for rational $\Theta$, unlike the case of general
$\Sigma$. We note also that the twisted Eguchi-Kawai model \cite{EK,GO} was
originally proposed as a model which reproduces large $N$ gauge theory in the
't Hooft limit (or the planar limit). In the present interpretation, however,
we have to take a different large $p$ limit, as we discussed in the previous
subsection, whereby there are non-planar contributions. Therefore, although the
matrix model is the same, the limiting procedure that one should take is
different in the two interpretations. Indeed, the scaling of Wilson loops in
such a non-planar large $p$ limit has been observed numerically
for the two-dimensional Eguchi-Kawai model in Ref.~\cite{NN}.

Let us also comment on the compatibility of finite noncommutativity parameters
and finite spacetime volume in the continuum limit. In Ref.~\cite{AIIKKT},
a particular choice of twist which corresponds to
$q_i = L^{d-1}$ in eq.~(\ref{Qdiag}) with $p = L^d$
was taken, where $D=2d$ is the dimension of spacetime. In that case, the
integers $\tilde p_i$ and $\tilde q_i$ defined in (\ref{pitildedefs})
are given by $\tilde p_i = L $ and $\tilde q_i = 1$.
One can then take $a_i = 0$ and $b_i = 1$ to satisfy (\ref{apbqpi}).
Therefore, the volume of the torus is $(\epsilon L)^D$
and the dimensionful noncommutativity parameters are given by
$\theta'_{\mu\nu}\propto L \epsilon ^2$.
As a result, in order to keep the noncommutativity parameters
finite in the continuum limit, one has to send the volume of the
discrete periodic lattice to infinity.\footnote{Indeed, in the two-dimensional
Eguchi-Kawai model studied in Ref.~\cite{NN}, the scaling of Wilson loops was
observed in the limit of large $L$ while keeping the parameter $L \epsilon ^2$
fixed.} In Ref.~\cite{AMNS}, this restriction was
removed by considering a restricted twisted Eguchi-Kawai model obtained
via the introduction of a quotient condition on the matrices which is the
discrete analog of that which appears in the Connes-Douglas-Schwarz formalism
\cite{CDS}. However, the present general construction suggests a much simpler
way
to avoid this constraint, namely by choosing the twist as described below
eq.~(\ref{Thetadiag}).

We close this Section by remarking that the above derivations are independent
of any representation of the operators $\hat{Z}_a'$ and $\hat{D}_\mu$. This
refines the approach of \cite{AMNS} whereby a specific finite dimensional
representation of these operators was used to interpret the twisted
Eguchi-Kawai model as noncommutative Yang-Mills theory. All we have used in the
above construction is the equivalence of commutative and noncommutative gauge
theories on the lattice at the field theoretical level, without ever having to
specify the
representation of the operators used. To this end, we note that the above
construction could have been obtained directly from the action (\ref{action})
by using the representation of the operators in terms of the $p \times p$
matrices
\beqa
\hat Z_a'&=&\prod_{b=1}^D(\Gamma_b)^{B_{ab}} \ , \\\hat D_\mu&=&\Gamma_\mu \ .
\eeqa
These matrices clearly satisfy the commutation relations
(\ref{Zprimelatticecomm}), and thereby constitute a finite dimensional
representation of the noncommutative torus.

\section{Coupling to fundamental matter fields}
\label{hopping}
\setcounter{equation}{0}

In this last Section we will consider noncommutative gauge theory coupled to
matter fields in the fundamental representation of the gauge group.
One advantage of the lattice formulation is that it allows
a hopping parameter (or large mass) expansion
of this theory which will enable us to clarify
various aspects of the Wilson loops in noncommutative gauge theories
that were constructed in Section~\ref{observables}.
A peculiar property of these Wilson loops,
in contrast to their commutative counterparts,
is that star-gauge invariance does not require that
the loops be closed, but it does require that the separation between their two
endpoints is proportional to the total momentum of the loop.
In the following we will clarify the physical meaning of these observables.
This analysis will also demonstrate explicitly how
these observables reduce smoothly to ordinary closed Wilson loops in the
commutative limit. For this, we will first show that the effective action for
the gauge field induced by integrating over the matter fields can be written as
a sum over all star-gauge invariant observables associated with closed loops.
We will then construct star-gauge invariant local operators out of the
matter fields and evaluate their correlation functions, thereby demonstrating
that the star-gauge invariant observables associated with open loops
appear very naturally in this way.

We will also show that Morita equivalence holds in noncommutative Yang-Mills
theory when it is coupled to fundamental matter fields. This Morita equivalence
clarifies the interpretation of the perturbative beta-function which was
calculated in Ref.~\cite{Hayakawa}. It also allows us to nonperturbatively
define noncommutative Yang-Mills theory with
fundamental matter fields, as we did in the matter-free case in
Section~\ref{simpleMorita}. As a special case, we obtain the twisted
Eguchi-Kawai model
with fundamental matter fields as constructed in Ref.~\cite{Das}.
The main idea of these constructions is to introduce a number of flavours $N_f$
which is equal to (or, more generally, a multiple of)
the number of colours, i.e. the rank of the gauge group.
This enables us to impose a boundary condition on the matter fields using a
rotation in flavour space to mimick the boundary condition
for adjoint representation fields. For simplicity, we will consider only
complex scalar matter fields. Fermionic matter fields can be treated in an
analogous way.

\subsection{Properties of star-gauge invariant observables}
\label{properties_hopping}

In this Subsection we will consider a discrete noncommutative torus and
periodic boundary conditions on all fields. We introduce the operator
\beq
\hat{\phi} =\sum_x\hat{\Delta} (x) \otimes \phi (x) \ ,
\eeq
where $\phi (x)$ is a complex scalar field
in the fundamental representation of the gauge group U($p$), and as always
the lattice operator $\hat{\Delta} (x)$ is defined in (\ref{map_lattice}).
The action for the matter field can be written as
\beq
S_{\rm matter} =- \kappa \sum _\mu \left\{
\Tr \left( \hat{\phi} ^\dag\,\hat{U}_\mu\,\hat{D}_\mu\,
\hat{\phi}\,\hat{D}_\mu^\dag\right)+ \mbox{c.c.} \right\}
+ \Tr \left( \hat{\phi} ^\dag\,\hat{\phi} \right) \ ,
\eeq
and it is invariant under the transformation
\beqa
\hat{\phi}&\mapsto&\hat{g}\,\hat{\phi} \ , \n
\hat{\phi} ^\dag&\mapsto&\hat{\phi} ^\dag\,\hat{g}^ \dag \ , \n
\hat{U}_\mu&\mapsto&\hat{g}\,\hat{U} _\mu\,
\hat{D}_\mu\,\hat{g} ^\dag\,\hat{D}_\mu ^\dag  \ .
\eeqa
In terms of lattice fields, the action can be written as
\beq
S_{\rm matter}= - \kappa   \left\{
\sum _{x,\mu} \phi(x) ^\dag  \star U_\mu (x) \star
\phi (x + \epsilon \hat{\mu})+ \mbox{c.c.} \right\}
+ \sum _x \phi(x) ^\dag\,\phi (x) \ ,
\label{matter_action}
\eeq
and it is invariant under the star-gauge transformation
\beqa
\phi (x)&\mapsto&g(x) \star \phi (x) \ , \n
\phi(x) ^\dag&\mapsto&\phi(x) ^\dag \star g(x)^ \dag \ , \n
U_\mu (x)&\mapsto&g (x) \star U_\mu (x) \star
g(x + \epsilon \hat{\mu}) ^\dag \ .
\label{gaugetr_matter}
\eeqa

When we integrate over the matter field $\phi (x)$, we make an expansion
with respect to the hopping parameter $\kappa$. This will require the
computation of various correlation functions of the scalar fields, which can be
calculated by using the formulae
\beqa
& &\sum _{x,y} \phi (x) ^\dag \star F(x)
\star\Bigl\langle \phi (x + v)\,\phi (y) ^\dag\Bigr\rangle _{\kappa = 0}
\star G(y) \star \phi (y + u)\n& &~~~~~~~~~~=
\sum _{x} \phi (x) ^\dag \star F(x)
\star G(x+v) \star \phi (x + v + u) \ ,
\label{formula1}
\eeqa
\beq
\sum _{x}\Bigl\langle \phi (x ) ^\dag  \star
F(x) \star \phi (x+v) ^\dag\Bigr\rangle _{\kappa = 0}
= \delta _{v,0}\,\sum _{x}  F(x) \ .
\label{formula2}
\eeq
Throughout this section, the brackets $\langle \cdots \rangle$ denote
the vacuum expectation value for fixed gauge background, i.e. we integrate over
the matter fields only. The suffix $\kappa =0$ means that the hopping parameter
is set to zero in the matter field action (\ref{matter_action}).
Star-gauge invariance requires that the lattice fields
$F(x)$ and $G(x)$ transform as
\beqa
F(x) &\mapsto & g(x) \star F(x) \star g(x+v) ^\dag \n
G(y) &\mapsto & g(y) \star G(y) \star g(y+u) ^\dag
\eeqa
under the star-gauge transformation (\ref{gaugetr_matter}).

Let us first consider the effective action $\Gamma _{\rm eff} [U]$
for the gauge field $U_\mu (x)$ induced by the integration over $\phi (x)$.
Using the hopping parameter expansion, it can be given as
\beq
\Gamma _{\rm eff} [U] = - \ln\left[\sum_{n=0}^\infty\frac{\kappa ^n}{n ! }
\,\left\langle\left(
\sum _{x,\mu} \phi(x) ^\dag \star U_\mu (x) \star
\phi (x + \epsilon \hat{\mu})+ \mbox{c.c.} \right)^n
\right\rangle _{\kappa = 0}\right] \ .
\eeq
Integrating over the matter field
using Wick's theorem and the formulae (\ref{formula1})
and (\ref{formula2}), we obtain
\beq
\Gamma _{\rm eff}[U]= \sum _C \frac{\kappa ^{L(C)}}{L(C)}\,\sum _x
\tr_{(p)}\Bigl( {\cal U}(x;C)\Bigr) \ ,
\eeq
where $\sum _C$ is the sum over all closed loops (with and without
self-intersection) on the lattice, and $L(C)$ denotes the number of links in
the contour $C$. In this way, we encounter the star-gauge invariant observables
associated with closed loops.

Let us now consider star-gauge invariant observables involving
matter fields such as
\beq
G_1 [f] = \left\langle \sum_ x\phi(x) ^\dag \star \phi (x)
\star f(x)\right\rangle \ ,
\label{G1fdef}\eeq
which is star-gauge invariant for arbitrary functions $f(x)$ on the lattice.
The lattice field $f(x)$ can be regarded as the wavefunction of the composite
operator $\phi(x) ^\dag \star \phi (x)$. We first perform a Fourier
transformation of $f(x)$ and express the observable (\ref{G1fdef}) as
\beq
G_1 [f] = \sum_{\vec k}\tilde{f} (k)\left\langle
\sum_x\phi(x) ^\dag \star \ee ^{i k_\mu x_\mu}\,\id_p\star \phi (x+v)
\right\rangle \ ,
\eeq
where
\beq
v_\mu = \theta _{\mu\nu} k_\nu \ .
\label{vmu}\eeq
Here we have used the fact that the plane wave $\ee ^{i k_\mu x_\mu}\,\id_p$
acts as a translation operator in the noncommutative field theory as explained
in
(\ref{Svprop})--(\ref{Sv}). Integrating over the matter field using the hopping
parameter expansion, Wick's theorem, and the formulae (\ref{formula1}) and
(\ref{formula2}), we arrive at
\beq
G_1 [f] = \sum _{\vec k}\tilde{f}(k)\,\sum _{C_v}\kappa ^{L(C_v)}\,\sum
_x\tr_{(p)}\Bigl( {\cal U}(x;C_v) \star \ee ^{i k_\mu x_\mu}\,\id_p\Bigr) \ ,
\label{G1ffinal}\eeq
where $\sum _{C_v}$ denotes the sum over all loops on the lattice
starting from the origin and ending at the lattice point (\ref{vmu}).
Thus we find the star-gauge invariant observables encountered in
Section~\ref{observables} associated with open loops. In the commutative case
$\theta _{\mu\nu} = 0$, the separation vector (\ref{vmu}) of the loops
vanishes independently of their momenta $k_\mu$.
Then the sum over loops in (\ref{G1ffinal}) contains only closed loops.
The sum over $\vec k$ can therefore be done explicitly
reproducing the wavefunction $f(x)$.
In the noncommutative case, however, the sum over loops depends
on $k_\mu$ and we cannot reverse the order of the first two sums in
(\ref{G1ffinal}). The separation vector $v_\mu$ of the two ends of the loop
$C_v$ grows with its momentum $k_\mu$.
This is a characteristic phenomenon in noncommutative field theories, and it is
another signal of the UV/IR mixing.
If one would like to have a higher resolution in one direction,
say in the $\mu = 1$ direction, by increasing the momentum $k_1$, then the
object will extend in the other directions proportionally to $\theta _{\mu 1}
k_1$. Thus the size of the object depends on its momentum.
On the other hand, if one considers the case where $\tilde{f}(k)$ is supported
on finite momenta $k_\mu$, then
one can take the $\theta _{\mu\nu} \rightarrow 0$ limit smoothly,
reproducing the commutative case.

Let us next consider a
two-point function of the composite operators which is defined as
\beq
G_2 [f,g] =\left\langle
\sum_ x\phi(x) ^\dag \star \phi (x) \star f(x)
\cdot \sum_ y\phi(y) ^\dag \star \phi (y) \star g(y)
\right\rangle  \ .
\eeq
The connected part of this correlation function can be readily computed as
before to give
\beqa
G^{\rm(conn)}_2 [f,g]&\defeq&G_2 [f,g] - G_1 [f]\,G_1 [g]\n
&=&\sum _ {\vec k,\vec p} \tilde{f}(k)\,\tilde{g}(p)
\,\sum _{C_u,C'_v} \kappa ^{L(C_u)+L(C'_v)}
\n& &\times\,\sum _{x,y}\tr_{(p)}\Bigl(  {\cal U}(x;C_u)
\star \ee ^{i p_\nu y_\nu}\,\id_p\star
{\cal U}(y;C'_v) \star \ee ^{i k_\mu x_\mu}\,\id_p\Bigr) \ ,
\label{G2}\eeqa
where the double sum over contours in (\ref{G2}) is restricted
in such a way that the parallel transport operator ${\cal U}(x;C_u)$
goes from $x$ to $y+u$ and ${\cal U}(y;C'_v)$ from $y$ to $x+v$.
The separation vectors $v_\mu$ and $u_\mu$ are related to the momenta
$k_\mu$ and $p_\mu$ respectively by $u_\mu = \theta_{\mu\nu} p_\nu$
and $v_\mu = \theta _{\mu\nu} k_\nu$. As with the one-point function
(\ref{G1ffinal}), the sums over $\vec k$ and $\vec p$ can be done explicitly in
the commutative limit, reproducing the wavefunctions $f(x)$ and $g(y)$.
In the noncommutative case, the sums over contours in (\ref{G2})
depend on the momenta $\vec k$ and $\vec p$ and one cannot interchange the
momentum and loop sums. Note that the two parallel transport operators in
(\ref{G2})  do not form a closed loop. The difference between the endpoint of
one and the starting point of the other are given by $u_\mu$ and $v_\mu$.
If one increases the momenta $k_\mu$ or $p_\mu$,
then the sizes $u_\mu$ or $v_\mu$ also increase.

The considerations in this section illustrate that the star-gauge
invariant observables constructed in Section~\ref{observables}
do indeed play the same fundamental role as ordinary Wilson loops do
in commutative gauge theories.
As the knowledge of all Wilson loop correlators would give
all the information about the quantum gauge theory coupled to matter fields,
so do the star-gauge invariant observables in the noncommutative case.
In particular, this result implies that the issue of whether or not a
noncommutative gauge theory is confining can be addressed by
searching for the area law behaviour of a star-gauge invariant observable
associated with a closed loop. We have also seen explicitly how these
observables reduce smoothly to ordinary Wilson loops in the commutative limit
for fixed gauge background.

\subsection{Morita equivalence with fundamental matter fields}
\label{Morita_fundamental}

In this section, we will describe Morita equivalence of noncommutative
Yang-Mills theories coupled to fundamental matter fields.
We consider the setup of Section \ref{simpleMorita}.
We start with {\em commutative} U($p$) gauge theory and
introduce $N_f$ flavours of
matter fields in the fundamental representation of the gauge group.
We represent it as a matrix $\Phi (x) _{iJ}$, where $i=1,\dots, p $ is the
colour index and $J= 1,\dots , N_f$ is the flavour index.
The spacetime is discretized as $x_\mu\in\epsilon\,\IZ$.
For the present purposes, it is essential to take $N_f$ to be
an integer multiple of $p$. In what follows, we will assume that $N_f = p$ for
simplicity. Then, $\Phi (x) _{iJ}$ becomes a $p \times p$ complex-valued
matrix, i.e. an element of ${\rm gl}(p,\IC)$.

The action for the gauge field is given by
(\ref{commutativeaction}). The action for the matter field is
\beq
S_{\rm matter}
= - \kappa
\left\{ \sum _{x , \mu} \tr_{(p)}\Bigl( \Phi(x) ^\dag \,  U_\mu (x) \,
\Phi (x + \epsilon \hat{\mu})\Bigr) + \mbox{c.c.} \right\}
 + \sum _x \tr_{(p)}\Bigl(\Phi (x) ^\dag\,\Phi (x)\Bigr) \ .
\label{matter_action_Morita}
\eeq
This matter-coupled gauge theory is invariant under the gauge transformation
\beqa
U_\mu (x)&\mapsto&g(x)\,U_\mu (x)\,g(x+\epsilon \hat{\mu})\n
\Phi (x)&\mapsto&g(x)\,\Phi (x)\n\Phi (x)^\dag&\mapsto&\Phi (x)^\dag\,
g(x)^\dag  \ .
\label{con_gaugetr_x_mat}
\eeqa
The action (\ref{matter_action_Morita}) also possesses
the global U($N_f$) flavour symmetry
\beq
\Phi (x) \mapsto   \Phi (x)\,g'~~~~~~;~~~~~~\Phi(x)^\dag\mapsto g'^\dag\,
\Phi(x)^\dag \ ,
\label{flavor_mat}
\eeq
where $g'\in \mbox{U}(N_f)$.

We impose twisted boundary conditions on the fields $\Phi (x)$ and $U_\mu (x)$
given by
\beqa
U_\mu (x+ \Sigma _{\nu a}\,\hat{\nu}) &=&
\Gamma _ a\,U_\mu (x)\,(\Gamma _ a)^\dag \n
\Phi  (x+ \Sigma _{\nu a}\,\hat{\nu}) &=&
\Gamma _ a\,\Phi  (x)\,(\Gamma_ a)^\dag \ ,
\label{Phi_constraint}
\eeqa
where $\Sigma$ is the period matrix and $\Gamma _ a $ are the twist-eaters
satisfying the commutation relations (\ref{GammaalgD_higher}).
Note that in (\ref{Phi_constraint}), the $\Gamma _ a$ acting on the left of
$\Phi (x)$ represent a (global) gauge transformation, whereas the
$(\Gamma _ a)^\dag$ acting on the right represent a rotation in flavour space.
This is the trick to introducing the fundamental representation of the gauge
group in Morita equivalence transformations.
Even though the matter fields define fundamental sections of the given gauge
bundle over the noncommutative torus (see (\ref{con_gaugetr_x_mat})), we can
exploit the global SU($N_f$) flavour symmetry to mimick
the boundary conditions for adjoint representation fields.
Such an idea first appeared in Ref.~\cite{CG}
in the context of supersymmetric field theories.

Rewriting the fields in terms of operators and solving the
constraints (\ref{Phi_constraint}) as we did in Section~\ref{simpleMorita},
we find that the resulting field theory is
noncommutative U($\tilde{p}_0$) lattice gauge theory,
on a lattice with period matrix $\Sigma ' = \Sigma\,\tilde{P} $
and with dimensionless noncommutativity parameters given by
$\Theta ' = - \tilde{P} ^{-1} B^\top $,
coupled to $\tilde{p}_0$ flavours of matter fields in the fundamental
representation of U($\tilde{p}_0$).
When $N_f = n_f\,p$, we obtain $n_f\,\tilde{p}_0$ flavours in the
Morita equivalent noncommutative field theory.
The one-loop beta-function of this theory for $\tilde{p}_0 = 1$ has been
calculated in Ref.~\cite{Hayakawa}. The contribution of fundamental matter
fields to the one-loop beta-function of noncommutative U($\tilde{p}_0$)
Yang-Mills theory with $n_f$ flavours coincides up to a finite rescaling of the
Yang-Mills coupling constant with that in commutative U(${p}$) Yang-Mills
theory with $N_f = n_f\,p$ flavours, since one-loop Feynman diagrams
involving fundamental matter fields are always planar. This coincidence
can now be understood as a manifestation at the one-loop level of
the Morita equivalence we have found in this Section.

As we did in Section \ref{TEK_model},
we can set $\Sigma = \epsilon \id _D$
to obtain the twisted Eguchi-Kawai model coupled to matter fields.
This is the model which was introduced in Ref.~\cite{Das} as a model
which reproduces large $N$ gauge theory with $N_f$ flavours of matter in the
Veneziano limit $N_f \sim N \rightarrow \infty$.
Here we have found that the same model with $N_f = n_fN$ can be interepreted as
noncommutative U($\tilde{p}_0$) gauge theory
with periodic fields including $n_f\,\tilde{p}_0$ flavours of matter fields in
the fundamental representation of the gauge group U($\tilde p_0$).

\subsection*{Acknowledgements}

The work of J.A. and Y.M. is supported in part by
MaPhySto founded by the Danish National Research Foundation.
Y.M. is sponsored in part by the Danish National Bank.
J.N. is supported by the Japan Society for the Promotion of
Science as a Research Fellow Abroad. The work of R.J.S. is supported in part by
the Danish Natural Science Research Council.

\end{document}